\documentclass[reqno,11pt]{article}

 \usepackage{amsmath}
 \usepackage{amsfonts}
 \usepackage{amssymb}
 \usepackage{amsthm}

 \usepackage{setspace}
 \usepackage{fullpage}

\usepackage{epsfig}
\usepackage{wrapfig}

\usepackage[]{graphicx}
\usepackage{epsfig}

 \newcommand{\hide}[1]{}

\newcommand{\Xomit}[1]{}

 \def\qed{\ifmmode$\blacksquare$\else{\unskip\nobreak\hfil
 \penalty50\hskip1em\null\nobreak\hfil$\blacksquare$
 \parfillskip=0pt\finalhyphendemerits=0\endgraf}\fi\vspace{0.3cm}}

 \newtheorem{theorem}{Theorem}[section]
 \newtheorem{observation}{Observation}[section]
 \newtheorem{definition}{Definition}[section]
 \newtheorem{lemma}{Lemma}[section]
 \newtheorem{corollary}{Corollary}[section]

\newtheorem{property}{Property}[section]
\newtheorem{remark}{Remark}[section]

 \newcommand{\vone}{\vspace{.1in}}
 \newcommand{\vhalf}{\vspace{.05in}}

 \newcommand{\ceil}[1]{\left\lceil{#1}\right\rceil}

\begin{document}

\title{\Large {Analysis of Randomized Work Stealing with False Sharing}
}
\author{Richard Cole ~\thanks{Computer Science Dept., Courant Institute
of Mathematical Sciences, NYU, New York, NY 10012.
Email: {\tt cole@cs.nyu.edu}.
This work was supported in part by NSF Grant CCF-0830516.}
\and Vijaya Ramachandran~\thanks
{Dept. of Computer Science, University of Texas, Austin, TX 78712. Email:
 {\tt vlr@cs.utexas.edu}. This work was supported in
 part by NSF Grant CCF-0830737.}
 }

 \maketitle

\pagenumbering{arabic}
\thispagestyle{empty}

\begin{abstract}

This paper analyzes the cache miss cost of algorithms when scheduled
using randomized
work stealing (RWS) in a parallel environment,
taking into account the effects of false sharing.

First, prior analyses~\cite{ABB02} are extended to incorporate false sharing.
However, to control the possible delays due to false sharing,
some restrictions on the algorithms seem necessary.
Accordingly, the class of Hierarchical Tree algorithms is introduced
and their performance analyzed.

In addition, the paper analyzes
the performance of a subclass of the Hierarchical Tree Algorithms, called
HBP algorithms, when scheduled using RWS;
improved complexity bounds are obtained for this subclass.
This class was introduced in~\cite{CR10a} with
%\marginpar{RJC 3/10}
% RJC several changes marked by RJC 3/10
% RJC 3/10 as agreed
efficient resource oblivious computation
% a multicore environment
in mind.

%RJC 3/10
Finally, we note that in a scenario in which there is no false sharing
the results in this paper match prior bounds for cache misses
% when there is no false sharing
but with reduced assumptions, and in particular with no need for a bounding  concave function
for the cost of cache misses as in~\cite{FS06}.
This allows non-trivial cache miss bounds
% RJC 3/10
in this case
to be obtained for a larger class of algorithms.

 \end{abstract}

\section{Introduction}\label{ssec:intro}

Work-stealing is a longstanding technique for distributing work
among a collection of processors~\cite{BS81,H84,WK91,BL99}.
Work-stealing operates by  organizing work in tasks,
with each processor managing its currently assigned tasks.
Whenever a processor $p$ becomes idle, it selects another processor $q$ and is given
(it \emph{steals}) some of $q$'s available tasks.
A natural way of selecting  $q$ is for $p$ to choose it uniformly at random from among the other
processors.
To emphasize the inherent randomization in this process, we call it randomized
work stealing, RWS for short.
RWS has been widely implemented, including in Cilk~\cite{BJ+95}, Intel TBB~\cite{RVK08}
and KAAPI~\cite{GBP07}.
% RJC need to check KAAPI paper
This methodology is continuing to increase in
importance due to its applicability to
multicore computers in which each processor (or core) has a private cache.

RWS scheduling has been analyzed and shown to provide
provably good parallel
speed-up for a fairly general class of algorithms~\cite{BL99}.
Its cache overhead was considered
in~\cite{ABB02},  which gave some general bounds
on this overhead;  these bounds were improved in~\cite{FS06}
for a class of computations whose cache complexity
function can be bounded by a concave function of the operation count.
The bound in~\cite{ABB02} when applied to processing a list of tasks
was recently improved in~\cite{TGTRB10}.

These analyses assume that there is no false sharing.
False sharing can occur as a result of employing cache coherency protocols,
when data is moved to and from cache in blocks (cache lines)
comprising multiple words.
False sharing refers to two different processors seeking to access distinct
locations in the same block, and if one or both seeks to
perform a write, only one of them can access the block at a time.
This reduces the possible parallelism and potentially
increases algorithm runtime.

In this paper, we analyze the efficiency of algorithms when
scheduled using RWS, taking account of delays due to false sharing.
As in prior work, there are two parts to the analysis; bounding the number of
steals, and bounding the additional costs of stolen tasks, which depend in part on
the number of steals.
Accounting for 
false sharing (or more generally, block misses)
affects both parts.

To achieve high efficiency, we will need algorithms that organize their writes
in a way that minimizes interaction among the different processors
and hence among the tasks forming the algorithm.
We will characterize a class of such algorithms, termed Hierarchical Tree Algorithms,
and analyze their performance.

In addition, we give a more refined analysis for a subclass of these algorithms,
the \emph{Hierarchically Balanced Parallel} (HBP) algorithms, which were 
% designed to be efficient in an oblivious multicore setting~
introduced in~\cite{CR10a, CR10b}.
This class includes
standard divide and conquer algorithms such as matrix multiply
(used as a running example in this paper), FFT, a new sorting algorithm~\cite{CR10b, CR10b2},
and some list and graph algorithms~\cite{CR10a}.

\section{Computation Model}
\label{sec:comp-model}

We model a computation using a
%\marginpar{RJC 3/10}
% change following the SPAA referee
directed acyclic graph, or dag,
% computation dag
$D$
(good overviews can be found in~\cite{LP98, BlLe99}).
$D$ is restricted to being a series-parallel graph,
where each node in the graph corresponds to a size $O(1)$ computation.
Recall that a directed series-parallel graph has start and terminal nodes.
It is either a single node, or it is created from two series-parallel graphs,
$G_1$ and $G_2$, by one of:
\begin{description}
\item
[i.] Sequencing, where the terminal
node
of $G_1$ is connected to the start node of $G_2$.
\item
[ii.] A parallel construct, which introduces a new start node $s$ and a new terminal node
$t$. $s$ is connected to the start nodes for $G_1$ and $G_2$, and their
terminal nodes are connected to $t$.
\end{description}
$D$ supports multithreaded computation by enabling two threads to continue from
each node $s$ in (ii) above; these threads then recombine into a single thread at
the corresponding node $t$.
This multithreading corresponds to a fork-join in a parallel programming language.

We will be considering algorithms expressed in terms of tasks, a simple task
being a size $O(1)$ computation, and
more complex tasks being built either by sequencing, or by forking, often
expressed as recursive subproblems that can be executed in parallel.
Such algorithms map to series-parallel computation dags.

In RWS, each processor maintains a work queue, on which it stores tasks that can be
stolen. When a processor $C$ generates a new stealable task it adds it to the bottom of
its queue. If $C$ completes its current task, it retrieves the task $\tau$ from the bottom of its queue,
and begins executing $\tau$.
The steals, however, are taken from the top of the queue.

An idle processor $C'$ will pick a processor
$C''$ uniformly at random
and independently of other steals, and attempts to steal from the top of
$C''$'s task queue.  If the steal
fails (either because the task queue is empty, or because some other processor
was attempting the same steal, and succeeded)
then processor $C'$ continues trying to steal,
continuing until it succeeds.

\hide{
An idle processor $C'$ will pick another random processor
$C''$ (uniformly at random
and independent of other steals), and attempts to steal from the top of
its task queue.  If the steal
fails (either because the task queue is empty, or because some other processor
was attempting the same steal, and succeeded)
then processor $C'$ continues to try to steal,
continuing until it succeeds in stealing a task.
} % end hide

We consider a computing environment comprising a collection of $p$ processors,
each equipped with a local  memory or cache of size $M$.
There is also a common shared memory of unbounded size.
Data is transferred between the shared and local memories in size $B$ blocks (or cache lines).

%\marginpar{RJC 3/10}
%RJC 3/10 complete rewrite of paragraph
We are mainly interested in algorithms that are optimal both in
terms of their cache miss cost and their work,
and on the maximum parallelism that can be achieved given these desiderata.
In this paper, we delineate constraints on the algorithms
that enable cache efficiency and show how to analyze the cache and
block miss costs of such algorithms.
As we will see, these constraints are not onerous:
they are observed by a variety of standard parallel algorithms,
modulo at most small changes.

\hide{
We seek algorithms that are efficient, ideally optimal, both in terms
of their operation count or work, and the number of cache misses they incur.
Given these goals, we also seek to maximize the possible parallelism.
As is standard, we compare the
work and cache miss bounds
to the best bounds
achievable by sequential algorithms.
} % end hide

\subsection{Cache and Block Misses}
\label{sec:c-miss}

We distinguish between two types of cache-related costs,
as discussed in \cite{CR10a}.

The term \emph{cache miss} denotes a
read of a block from shared-memory into processor $C$'s cache,
when a needed data item is not currently in
the cache, either because the block was never read by processor $C$, or because
it was evicted from $C$'s cache to make room for new data.
This is the standard type of cache miss that occurs, and is
accounted for, in sequential cache complexity analysis.

%RJC new version
The term \emph{block miss} denotes an
update by a processor $C' \neq C$ to an entry in a block $\beta$ that is
in processor $C$'s cache.
This results in block $\beta$ being invalidated,
and results in processor $C$ needing to read in block $\beta$ the next
time it accesses data in this block. This is done
so that data consistency is maintained within the elements of a block
across all copies in caches at all times.
This type of `cache miss' does not occur in a sequential computation
and seems challenging to bound. We will refer to this
type of caching cost as a {\it block miss};
this includes the cost of `false sharing'.
There are other ways of dealing with block misses, but
we believe that the block miss cost with our invalidation rule
is likely as high as (or higher than) that
incurred by other mechanisms. Thus, our upper bounds should hold
for most of the coping mechanisms known for handling block misses.

\hide{
The term \emph{block miss} denotes an
update by a processor $C' \neq C$ to an entry in a block $\beta$ that is
in processor $C$'s cache.
This results in block $\beta$ being invalidated,
and results in processor $C$ needing to read in block $\beta$ the next
time it accesses data in this block.
This type of `cache miss' does not occur in a sequential computation.
and seems challenging to bound. We will refer to this
type of caching cost as a {\it block miss};
this includes the cost of `false sharing'.
} % end hide

Work stealing causes the algorithm execution to incur additional cache
misses and introduces block misses.
Our analysis does not make any assumptions about the mechanism used
for handling accesses to a shared block and in particular
does not assume it is fair.
This can result in block misses being unboundedly expensive.
Instead, we use algorithmic techniques to control these costs.

\hide{
Work stealing causes the algorithm execution to incur additional cache
misses and introduces block misses.
} % end hide

\subsection{The Main Results}
\label{sec:results}

To achieve good bounds, we consider  algorithms that exhibit good data locality.
Techniques enabling this have
been developed for cache aware and cache oblivious algorithms;
we will introduce additional methods to help control the cost of block misses.
In particular, we characterize a class of recursive algorithms
we call \emph{Hierarchical Tree Algorithms}, which, in addition, satisfy the following
properties:

i. Data locality in the writes.

ii. Limited access: each writable variable is written $O(1)$ times.

iii. A constraint on space usage, we call \emph{top-dominance}.

iv. A sufficient shrinkage in the size of recursive subproblems.

These
properties are also used in our companion paper~\cite{CR10a}
in the analysis of its deterministic PWS scheduler.

At this point, a definition of \emph{task size} will be helpful.

\begin{definition}
\label{def:task-size}
A task $\tau$ is said to have size $r$, denoted $|\tau|$, if it accesses $r$ distinct
\emph{(}one-word\emph{)} variables over the course of its execution.
\end{definition}

We use the following parameters to specify our results.
Let $\cal A$ be an algorithm described by a series-parallel dag $D$.
Suppose that on an input of size $n$, in the worst case, in a sequential execution,
$\cal A$ performs $W$ operations and incurs $Q$ cache misses.
Suppose that an operation on in-cache data takes $O(1)$ time units,
that the cost of a cache miss is $O(b)$ time units,
%VLR
% and 
%
the cost for stealing a task is $\Theta(s)$ time units,
%VLR
and the cost for an unsuccessful steal is $O(s)$, allowing for the
possibility that an unsuccessful steal has less cost than a 
successful one.
% We also suppose that the cost for an unsuccessful steal is $O(s)$, by
% contrast with previous work that assumed a cost of $\Theta(s)$.
%
Let $T_{\infty}$ be the maximum length of the paths descending the dag $D$
representing $\cal A$'s computation,
let $E$ be a bound on the cost, measured in
cache misses, incurred in performing reads and writes at any single node of $D$,
and let $D_b$ be a bound on the cost, measured in
cache misses, incurred in performing reads and writes on any single path in any execution of $D$.
Clearly, $D_b \le E T_{\infty}$.

\vhalf
\noindent
The analysis has four parts:

\vhalf
\noindent
1. A bound on the cost of cache misses as a function of the
number $S$ of stolen tasks. We obtain the same bounds as Frigo and Strumpen~\cite{FS06},
but with a more direct analysis that avoids the need for an assumption that the cost of
cache misses is bounded by a concave function of the work performed by a subtask.
Rather, we use an assumption that the cost of
cache misses is bounded by a non-decreasing function of the size of a task.

We  illustrate this methodology by example,
using two standard algorithms for matrix multiply.
The first algorithm is the depth $\log^2 n$ recursive algorithm that calls 8 subproblems in parallel.
The second algorithm is the depth $n$ in-place algorithm,
which calls two sets of 4 subproblems in sequence;
this algorithm does not observe
the limited access property (from (ii) above); we make a small modification,
which ensures the property holds (but the algorithm is no longer in-place).

This methodology can handle algorithms not covered by the 
%VLR
%prior method,
the method in \cite{FS06},
which appears to mainly handle in-place recursive algorithms.
By contrast, many of the algorithms the new analysis is applied to use
recursive procedures that also call a distinct second recursive procedure
(e.g.~a matrix addition inside a matrix multiply).

Since the bit-interleaved (BI) format is more cache-efficient than
the Row Major (RM) format for matrix algorithms, we analyze both
matrix multiply algorithms assuming the BI format.
Accordingly, we also describe algorithms for converting between these
two formats.
The natural recursive algorithm for converting RM to BI is optimal both
in terms of its operation and its cache and block
miss cost; however, this is not true for the natural algorithm for
converting BI to RM, which potentially has a high cost due to block misses.
Instead, in Section~\ref{sec:alg-ex-bl-miss},
we will describe a slower algorithm for performing
this conversion, but one that is more efficient
in terms of its block misses.
The costs of this algorithm are dominated by those for the
matrix multiply algorithms
we consider, and consequently, the complexities of these
matrix multiply algorithms
are the same whether
their inputs and outputs are in RM or BI format.
%\marginpar{RJC 3/10}
% RJC forward pointer as you suggested
Another, yet better conversion algorithm 
%VLR
from \cite{CR10a} 
is mentioned briefly
in Section~\ref{sec:alg-bounds}.

\vhalf
\noindent
2. A bound on the additional cost due to block misses. This will be $O(B)$ per stolen
task for the class of algorithms we consider. This will follow from the above properties:
Properties (ii)-(iv) allow us to show
an $O(B)$ bound on the delay due to block misses
in accessing any one shared block.
Property (i) is concerned with limiting the number of shared
blocks per task; it is
an $O(1)$ bound
for the algorithms we consider and is an individual algorithm design issue.

\vhalf
\noindent
3. A bound on the number of successful steals,
which applies to \emph{any} computation described by a series-parallel computation dag,
and not just those covered by part (2) above,
so long as there is a bound $E$ on the  cost of cache and block misses as specified above.
This uses an analysis similar to that of Acar et al.~\cite{ABB02},
but will depend in part on the cost of the block misses, and also loosens 
the assumption
regarding the cost of an unsuccessful steal
%VLR
by allowing an unsuccessful steal to cost less than a successful one.
We have:

\vhalf
\noindent
With probability $1 - 2^{-aT_{\infty}}$,
for $a = \omega(1)$,
%VLR
a series-parallel computation
$\cal A$ has
at most the following number of successful steals:
\[
O\left( p\left[ T_{\infty} + \frac{b}{s} E T_{\infty}  \right](1 + a)
 \right).
\]
For the class of algorithms we consider, by (2), $E=O(B)$.
%\marginpar{RJC 3/10}
% RJC added sentence
For this class of algorithms,
$D_b = O(E T_{\infty})$ but it is not clear whether this is a tight bound.

\vhalf

\noindent
4. An improved bound on the number of successful steals for the case of
HBP algorithms~defined in \cite{CR10a}.
These algorithms require forked subproblems to be of roughly equal size. They form a
subclass of the Hierarchical Tree Algorithms.
For this class of algorithms, the impact of the block misses on the number of
steals can be bounded more sharply.
% RJC 3/10 reworded
% Compared to the result in (3), we
We
improve the bound in (3), replacing the term $\frac{b}{s}  E T_{\infty}$
by $\frac{b}{s} D_b$.
%\marginpar{RJC 3/10}
% have the following improved bound:
%
\hide{
\[
O\left(
 p \left[T_{\infty} + \frac{bD_b}{s}  \right] (1 + a)
\right).
\]
} % end hide
This is significant for we are able to obtain and use tighter bounds
% RJC 3/10
on
% for
$D_b$ than the
% RJC 3/10
earlier
% simple
$O(E T_{\infty})$.
%  mentioned earlier.
%
In particular,
for a computation described by an $m$-leaf forking tree,
$D_b$ reduces from $O(B\log m)$ to $O(\min\{B, m\} + \log m)$.
Corresponding improvements are obtained for the $D_b$ terms for more complex algorithms.
This result is shown in Theorems~\ref{thm:rws-work} and~\ref{th:path-length}.

\vhalf

Finally, in Section~\ref{sec:alg-bounds},
we derive complexity bounds for several other HBP algorithms.

\vspace*{-0.1in}

\section{Bounding the Cache Misses}
\label{sec:c-miss-anal}
%

%\marginpar{RJC 3/10}
% RJC new paragraph
Our method for bounding the cache misses determines bounds on the worst case number
of tasks of a given size that can be stolen in any execution.
For stolen tasks of size $2M$ or larger, up to constant factors, the same
cache miss costs would be incurred even if there were no steal.
For smaller tasks, the incurred costs are a function of the task size,
which combined with the bounds on the number of tasks of a given size, yields
bounds on the cache miss costs as a function of the number of stolen tasks.

We illustrate this method
\hide{
how to show bounds on the cache misses
as a function of the number of steals,
}% end hide
using two algorithms for matrix multiply (MM).
We assume that the matrices are in
the {\it bit interleaved \emph{(}BI\emph{)} layout}, which recursively
places the elements in the top-left quadrant, followed by the
recursive placement of the top-right,
bottom-left, and bottom-right quadrants.
This layout is well-known to be the effective in minimizing data
%\marginpar{RJC 3/10}
% RJC responding to referee request for citation
movement~\cite{FLPR99-short}.
In Section~\ref{sec:alg-ex-bl-miss},
we discuss algorithms for converting between the row major (RM)
and BI layouts.

\paragraph{Depth $\mathbf{n}$ MM}
The algorithm we consider, from~\cite{FLPR99}, recursively multiplies an initial four pairs of
$n/2 \times n/2$ matrices, recording the results, and then multiplies a second
collection of four pairs, adding the second set of results to the results from the initial multiplications.
It has $W = O(n^3)$,
$Q = O(n^3/(B\sqrt M))$, and $T_{\infty}= O(n)$.

We now demonstrate the same cache miss bound as obtained by
Frigo and Strumpen~\cite{FS06}.
But this is a slightly different algorithm as Frigo and Strumpen have
the matrices in RM
format (we need the BI format to control the block miss costs as
we will see later).

\begin{lemma}
\label{lem:depth-n-MM-cache-miss}
The depth $n$ MM algorithm incurs
$O(n^3/(BM^{1/2}) + S^{1/3}\frac{n^2}{B} + S)$
cache misses when it undergoes $S$ steals.
\end{lemma}
\begin{proof}
We begin by bounding the
cache miss cost for  the initial task and the stolen tasks of size $2M$ or larger;
we will show that collectively they incur
$O(n^3/(B\sqrt{M}))$ cache misses.
For simplicity, we assume that $n$ and $M$ are integer powers of 2.
Ignoring the work removed by smaller stolen subtasks, these tasks each solve one or more distinct
matrix multiply subproblems of size $2M$.
Each such subproblem incurs $O(M/B)$ cache misses, and there are $\Theta(n^3/M^{3/2})$
of them.
This yields a total of $O(n^3/(BM^{1/2}))$ cache misses incurred by these tasks.

For smaller stolen tasks $\tau$, the cache miss cost is $O(\ceil{|\tau|/B)})$.

In fact, the thread for a stolen task can exhibit an expanding size as it proceeds.
For example, a steal could be of say a $1\times 1$ MM subproblem,
which because it is the last subproblem to complete at the join, continues to execute
the remaining half of a $2\times 2$ MM subproblem (comprising the remaining four
$1\times 1$ subproblems), which is followed by the remaining half of a $4\times 4$ MM subproblem,
etc. However, for this example, and indeed for every steal, the data needed for the last (and largest) remaining half MM
subproblem computed, $\tau$ say, contains the data needed for all the earlier subproblems the steal computes,
and therefore if this last subproblem is of size $M$ or smaller, the steal incurs $O(\ceil{|\tau|/B)})$ cache
misses. Consequently, the prior analysis continues to apply.

It follows that the worst case bounds arise assuming the small stolen tasks are all as large as possible.
For each integer $i\ge 0$,
there are $O\left( 8^i \left( \frac{n}{\sqrt{M}}\right)^3 \right)$ stolen subtasks of size $M/4^i$.
They cause $O\left(  \left(2^i\frac{M}{B} + 8^i\right)\cdot \left( \frac{n}{\sqrt{M}}\right)^3  \right)$
cache misses.
Summing over all $i$, given that there are $S$ stolen tasks in all,
yields that $S = O\left( 8^i \left( \frac{n}{\sqrt{M}}\right)^3 \right)$ and hence gives
a total of $O(S^{1/3}\frac{n^2}{B} + S)$ cache misses.
\end{proof}
We will return to this bound
later when we have obtained bounds on $S$.

The standard way of implementing this algorithm is as an in-place process.
However, this violates the limited access property (for each output array location is
written $n$ times).
We make the algorithm limited access as follows:
for each recursive subproblem, we create a local array to store the results of its subproblems,
which are then added together and written to the array for the parent subproblem.
This increases the number of operations by a factor of 2, but this can be reduced to, for example,
less than 1\%
additional operations by having the base case comprise $10 \times 10$ matrices.
The local arrays also increases the space usage to $O(n^2\log p)$.
This appears to be a necessary part of our method for controlling block misses.

Finally, we note that the cache miss analysis of Frigo and Strumpen cannot be immediately
applied to this variant of MM. The reason is that there are subtasks which are matrix additions,
and they have a larger cache miss to work ratio than
the matrix multiply tasks. By contrast, the cache miss to size ratio is only reduced, and this
immediately yields the bound of Lemma~\ref{lem:depth-n-MM-cache-miss} for this variant of the algorithm.

\begin{corollary}
\label{cor:depth-n-MM-cache-miss}
The limited access version of the depth $n$ MM algorithm incurs
$O(n^3/(BM^{1/2}) + S^{1/3}\frac{n^2}{B} + S)$
cache misses when it undergoes $S$ steals.
\end{corollary}
\begin{proof}
The number of addition subtasks of
a given size, up to constant factors, is no larger than the number of multiplication subtasks
of the same size.
Thus the presence of these addition subtasks, which are no more expensive in
cache misses than the multiplication subtasks, does not alter the prior bound on cache misses.
\end{proof}

\paragraph{Depth $\mathbf{\log^2 n}$ MM}
This algorithm
multiplies two $n\times n$ matrices by recursively multiplying eight
$n/2 \times n/2$ matrices, followed by a tree computation
to add pairs of these recursively multiplied matrices.
This algorithm has $T_{\infty}=O(\log^2 n)$, $W=O(n^3)$,
and $Q = O(n^3/(B\sqrt M))$.

The exact same analysis applies to this algorithm, yielding:
\begin{corollary}
\label{cor:depth-log-2-n-MM-cache-miss}
The depth $\log^2 n$ MM algorithm incurs
$O(n^3/(BM^{1/2}) + S^{1/3}\frac{n^2}{B} + S)$
cache misses when it undergoes $S$ steals.
\end{corollary}
The difference, as we will see, is that this algorithm incurs far fewer steals than the depth $n$
algorithm.

\paragraph{Space Usage}
The depth $\log^2 n$ MM  algorithm uses space $O(p^{1/3}n^2)$, which is larger than the $O(n^2 \log p)$
space usage of the limited access depth $n$ MM algorithm,
which in turn is larger than the in-place $O(n^2)$ space use of the depth $n$ algorithm,
but for this latter algorithm
it is not clear whether there are
good bounds on the block delay costs.

\section{Bounding Block Misses}
\label{sec:lim-acc}

The delay caused by different processors
writing into the same block can be quite significant, and this is a
caching delay that is present only in the parallel context.
These costs might arise if two processors are
sharing a block (which occurs for example if data partitioning does not
match block boundaries) or
if many processors access a single block
(which could occur if the processors are all executing very small tasks).
We refer to any read of a block that is not in cache due to the block being
shared by multiple processors as a block miss.

In particular,
consider a parallel execution in which two or more processors between
them perform multiple
accesses to a block $\beta$, which include
$x \geq 1$ writes.
These accesses could cause
$\Theta (b \cdot x)$ delay at {\it every} processor accessing $\beta$, where
$b$ is the delay due to a single cache miss.
We will measure this delay in units of size $b$, the bound on the cost of a cache miss.
Consequently, henceforth, we will refer to this as a $\Theta(x)$ block delay,
which we associate both with the block and the processors accessing the block.

Further, $x$ can be arbitrarily large unless
care is taken in the algorithm design.
We establish that $x=O(B)$ when our recursive algorithms
use limited access variables (Property~\ref{def:limited-use-var})
and are exactly linear space bounded (a special case of Property~\ref{prop:space-use}),
and the runtime system observes a natural space allocation property
(Property~\ref{prop:sp-alloc}).
Of course, it is only when there are one or more writes to a block
that there can be any block misses.

\begin{definition}
\label{def:block-delay}
Suppose that block $\beta$ is moved $m$ times
from one cache to another \emph{(}due to cache or block misses\emph{)}
during a time interval $T = [t_1,t_2]$.
Then $m$ is defined to be the\emph{ block delay} incurred by $\beta$ during $T$.

The \emph{block wait cost} incurred by a task $\tau$ on a block $\beta$
is the delay incurred during the execution of $\tau$
due to block misses when accessing $\beta$, measured
in units of cache misses.
\end{definition}

If a task $\tau$ executes during time interval $T_{\tau}$ and accesses
block $\beta$ during its execution, then clearly the block wait cost
incurred by $\tau$ on $\beta$ is no more than the
block delay of $\beta$ during $T_{\tau}$.

\vhalf
In order to analyze the block miss
costs, we need to explain
how the program variables are stored.
Let $\tau$ be either the original task in the computation of
$\cal A$ or a stolen subtask.
When a task $\tau$ is initiated,
an execution stack
$S_{\tau}$ is created to keep track of the procedure calls and variables
used in $\tau$'s execution.
While this execution stack is created by the processor $C$
that starts $\tau$'s execution,
if another processor $C'$ takes over $\tau$'s execution, by being the second (and hence last)
processor to finish the work preceding a join, then $C'$ will continue using $S_{\tau}$
for the remainder of $\tau$'s computation (at least up until yet another processor takes
over $\tau$'s computation).
The variables on $S_{\tau}$ may be accessed by stolen subtasks also.
As $S_{\tau}$ grows and shrinks, with each growth period corresponding to the creation of
new variables, these new variables may be stored in reused
portions of a block $\beta$,
and this may happen repeatedly.
In Lemma \ref{lem:limited-write-intf-first} we show a bound of
$Y(|\tau|, B)$, for a suitable function $Y$,
on the block delay $\tau$ faces in accessing
a block $\beta$ on $S_{\tau}$.
For the algorithms we consider, $Y(|\tau|, B) = \min\{|\tau|, B\}$.

\hide{
In order to analyze the block miss
costs, we need to explain
how the program variables are stored.
When a processor $C$ steals a task $\tau$, it will create an execution stack
$S_{\tau}$ to keep track of the procedure calls and variables
in the work it performs on $\tau$.
The variables on $S_{\tau}$ may be accessed by stolen subtasks also.
As $S_{\tau}$ grows and shrinks it may use and then stop using portions of a block $\beta$
repeatedly.
In Lemma \ref{lem:limited-write-intf-first} we show a bound of
$Y(|\tau|, B)$, for a suitable function $Y$,
on the block delay $\tau$ faces in accessing
a block $\beta$ on $S_{\tau}$.
For the algorithms we consider, $Y(|\tau|, B) = \min\{|\tau|, B\}$.
} % end hide

In addition to the variables stored on the execution stacks, the
algorithm needs variables in which to
store its output.
These output variables, which may be arrays, are stored in memory locations
separate from
those used for the execution stacks, and
share no blocks
with the execution stacks.

At this point, it will be helpful to define the notions of local and global variables with respect to
a procedure $P$ of an algorithm.

\begin{definition}
\label{def:glob-loc-var}
A variable $x$ declared in a procedure $P$ is called a \emph{local} variable
of $P$.
A variable $y$ accessed by $P$ and declared in a procedure $Q$ calling $P$
or
%\marginpar{RJC 3/7}
% RJC I prefer for
% \marginpar{VLR:3/2}
%VLR: Removed the `for' (or did you want to remove the `in'?)
used
%RJC
for
% in
%
the inputs or outputs of the
algorithm $\cal A$ containing $P$ is said to be \emph{global} with
respect to $P$.
However, note that $y$ would be a local variable of $Q$ if declared in $Q$.
\end{definition}

\subsection{Algorithmic Constraints}
 \label{sec:alg-cstrnts}

The following lemma will motivate our definition of limited access algorithms.

\begin{lemma}
\label{lem:io-block-miss-bdd}
Let $\tau$ be a task whose execution is initiated by processor $C$, and
let $\beta$ be a block provided to $C$ to store local
variables of $\tau$.
Let $T$ be the time interval during which $\tau$ is executed, and let $T'$
be a subinterval of $T$ (possibly $T'=T$).
Suppose that processors $C_1, \cdots, C_k$ are the only processors executing stolen
subtasks of $\tau$ during $T'$.
Further suppose that they access block
$\beta$ a total of $x$ times during $T'$.
Then $\beta$ incurs a block delay of at most $2x$ during $T'$.
\end{lemma}

\begin{proof}
Processors $C_1,\cdots, C_k$ cause at most $x$
moves of block $\beta$ to their caches
as a result of their $x$ accesses.
Thus processor $C$ needs at most $x$
moves of block $\beta$ to its cache to handle
all its accesses, regardless of their number.
\end{proof}

We begin by specifying the notion of an \emph{access bound} for a task.

\begin{definition}
Let $\tau$ be a task and $\beta$ a block on the execution stack of the processor $C$
executing $\tau$.
Let $\tau_1, \cdots, \tau_k$ be the subtasks of $\tau$ stolen from $C$.
$\tau$ is defined to have access bound $s$ if $\tau_1, \cdots, \tau_k$
access $\beta$ at most $s$ times, for every such block $\beta$.
\end{definition}

Lemma~\ref{lem:task-bound} is an immediate consequence of Lemma~\ref{lem:io-block-miss-bdd}.

\begin{lemma}
\label{lem:task-bound}
If
stolen subtask
$\tau$ is $s$ access bounded,
then each block $\beta$ on $S_{\tau}$ incurs an $O(s)$ block delay.
\end{lemma}

We now identify a class of algorithms for which the access bound for their
stolen subtasks
$\tau$ is
$O(\min\{|(\tau|, B)\}$.

Our first constraint places a limit on how often each writable variable can be accessed.
\begin{property}
\label{def:limited-use-var}
An algorithm is \emph{limited-access} if each of its writable
variables is accessed $O(1)$ times.
\end{property}
Due to procedure calls, including recursive ones,
over time more than $B$ variables could all share a single block, and so
this property does not suffice to yield an $O(B)$ bound on the number of
accesses to a single block.

\vone
Our second constraint
imposes upper and lower bounds on the space used by the tasks
in our algorithms.
To help specify this we create the following hierarchy of algorithms.

\begin{definition}
A  \emph{Tree Algorithm} $\cal A$ is formed
from the down-pass of a binary forking computation tree $T$
followed by its up-pass, and satisfies the following additional properties.

\vhalf
\noindent
1. Each leaf node performs $O(1)$ computation.
Each non-leaf node in the down-pass performs only $O(1)$ computation before it forks its two children.
Likewise, each non-leaf node in the up-pass performs only $O(1)$ computation after the completion of its forked subtasks.

\vhalf
\noindent
2. Each node declares at most $O(1)$ local variables.
In addition, $\cal A$ may use size $O(|T|)$ global arrays to store its output.
If $\cal A$ is being used as a subroutine, these arrays are declared by the calling
procedure; otherwise, they are the output arrays for the algorithm.
\end{definition}

% \marginpar{VLR:2/24}
%
%VLR: This is all crystal clear now. The only qustion is whether
% we have talked too much about this -- in Defn 4.2 and the text above it,
% and then again here, and also further on below.
% But we can re-visit this at the end when the paper
% is about to be finalized.
%
% RJC: I too wondered if I was overdoing it.
%
Note that if $\cal A$ is a subroutine, then the global array for its output is stored
on the execution stack of the procedure that calls $\cal A$. While if it is the full algorithm,
its global array is the algorithm output, which is stored separately from any
execution stack.

Thus, on its execution stack, the task for a Tree Algorithm with computation tree $T$ will use space proportional
to the height of $T$.
Typically, in an efficient parallel algorithm, this height is $O(\log |T|)$.
Henceforth, for short, we  call such a task a \emph{tree task}.
In addition, the task initiating the computation will use space $O(T)$
for the global arrays on its execution stack.

We create more complex algorithms, which we call \emph{Hierarchical Tree Algorithms}, using sequencing
and recursion.

\begin{definition}
\label{def:rec-alg}
A \emph{Hierarchical Tree Algorithm} is one of the following:

\vhalf
\noindent
1.  A Type 0 Algorithm, a sequential computation of constant size.

\vhalf
\noindent
2. A Type 1, or Tree Algorithm.

\vhalf
\noindent
3. A Type $i+1$ Hierarchical Tree Algorithm, for $i \geq 1$.
 An algorithm $\cal{A}$ is a Type $i+1$ Hierarchical Tree Algorithm if,
on an input of size $n$, it calls, in succession,
a sequence of $c \ge 1$ collections of $v(n) \ge 1$ parallel recursive subproblems,
where each subproblem has size $s(n) \le n/b(n)$,
with $b(n) \ge 1+\nu$ for some constant $\nu > 0$;
further, each of these collections
can be preceded and/or followed by
$O(1)$ calls to Hierarchical Tree Algorithms of type at most $i$.

Data is transferred to and from the recursive subproblems by means of variables (arrays) declared at
the start of the calling procedure.
Note that these arrays are local to the calling procedures, and
global w.r.t.~the recursive procedures.

\vhalf
\noindent
4. A type $\max\{t_1, t_2\}$ Hierarchical Tree Algorithm results if it
is a sequence of two Hierarchical Tree Algorithms of types $t_1$ and $t_2$.
\end{definition}

The recursive forking of $v(n)$ parallel tasks in a Hierarchical Tree Algorithm
is incorporated into the binary forking in our multithreaded set-up by using a fork-join
structure identical to that for the tree algorithms, except that
each leaf of this tree corresponds to a recursive subproblem.

\subsubsection{Space Constraints}

We begin with two definitions regarding the space usage.

\begin{definition}
\label{space-usage}
An algorithm $\cal{A}$ with an input of size $n$ is \emph{Exactly
$S^l(n)$ Space Bounded}, also denoted by $S^l(\cal{A})$, if it stores its local variables in space of size between $S^l(n)$ and
$d S^l(n)$, for some constant $d \ge 1$.
If $\cal{A}$ is a Type $i$ Hierarchical Tree Algorithm, $i \ge 2$,
in which each recursive call has an input of size at most
%\marginpar{RJC 3/7}
% RJC changed
%VLR: Eveywhere else we use $s(n) \leq n/b(n)$,  why use $s(n) < n$ here?
$s(n)<n/b(n)$, where $b(n) \le 1 -\nu$ for some constant $\nu$, $0\le \nu \le 1$,
then $\cal{A}$ is defined to have \emph{Path Space Bound} $S^p(\cal{A})$ or
$S^p(n) =\sum_{i \ge 0} c^i S^l( s^{(i)}[n] )$,
where $c$ is the number of collections of recursive calls it makes.
If $\cal{A}$ is obtained by sequencing Hierarchical Tree Algorithms
${\cal{B}}$ and ${\cal{C}}$,
then $S^p(\cal{A}) = \max\{$$ {S^p}( {\cal{B}} )$, ${S^{p}} ( {\cal{C}} )$$ \}$.
While if $\cal{A}$ is a tree computation $T$, then
$S^p({\cal{A}}) = S^p(n)=\Theta(\operatorname{height}(T))$.
\end{definition}

The following
Property~\ref{prop:space-use} imposes a \emph{lower} bound on $S^l(\cal{A})$.
This property will be used in the proof of Lemma~\ref{lem:limited-write-intf-first}.

\begin{property}
\label{prop:space-use}
A Type $i+1$ Hierarchical Tree Algorithm $\cal{A}$, $i\ge 1$, is \emph{top-dominant}
if for each type $h$ procedure $\cal{B}$
it calls, $h\le i$,
$S^l({\cal{A}})=\Omega( S^{p} ( {\cal{B}} ))$, and for $h\ge 2$,
$\cal{B}$ is also top-dominant.
\end{property}
\noindent
In all the recursive algorithms we consider,
$S^l(n), S^p(n)=\Theta(n)$, and the tree algorithms have $S^l(n) =O(1)$ and $S^p(n)=O(\log n)$.
Thus these algorithms are all top-dominant.
We also say such algorithms are
\emph{Exactly Linear Space Bounded}.

The second space constraint concerns space allocation by the runtime system.

\begin{property}\label{prop:sp-alloc}
\emph{(Space Allocation Property.)}~~Whenever a processor requests space
it is allocated in block sized units; naturally, the
allocations to different processors are disjoint and entail no block sharing.
\end{property}

\subsection{Analysis of the Cost of Block Misses}

\begin{definition}
\label{def:usurpation}
The \emph{kernel}
of $\tau$ is the portion of $\tau$ remaining after its stolen subtasks are removed.

We note that the processor $C$ executing $\tau$'s kernel
may change after a join of subtask $\tau''$ executed by $C$
with stolen subtask $\tau'$ executed by $C'$, in the event that
$C'$ finishes executing $\tau'$ after $C$ finishes its execution of $\tau''$;
then $C'$ will take over the remainder of the
execution of $\tau$'s kernel.
If there is such a change, we call it a \emph{usurpation} by $C'$.
\end{definition}

The following observation is readily seen and will be used in the proof of Lemma~\ref{lem:BP-block-miss}.
\begin{observation}
\label{obs:stolen-tasks}
Let $D$ be the series-parallel computation dag for a
task $\tau$.
Let $v$ be the node in $D$ corresponding to the last task $\tau_v$
to be stolen during the execution of  $\tau$'s kernel,
and let
$P_{\tau}$
be the path in $D$ from the root of $D$
to the parent of $v$.
Then, the set of tasks stolen from the processor(s) executing $\tau$'s kernel
consists of some or all of the tasks corresponding to those nodes of $D$ that
are the right child of a node on
$P_{\tau}$
but are not themselves on
$P_{\tau}$.
Further, they are stolen in top-down order with respect to the path
$P_{\tau}$.
\end{observation}

\begin{lemma}
\label{lem:BP-block-miss}
Let $\cal A$ be a limited-access Tree Algorithm and let
$\tau$ be either the original task in the computation of
$\cal A$ or a task which is stolen during the
execution of $\cal A$.
Let $\beta$
be a block used for
$\tau$'s execution stack $S_{\tau}$.
Then $\beta$ incurs a block delay of $O(\min\{B, \operatorname{ht}(\tau)\})$ during  $\tau$'s execution,
where $\operatorname{ht}(\tau)$ denotes the height of the corresponding computation tree.
\end{lemma}
\begin{proof}
By Observation \ref{obs:stolen-tasks},
there is a single path $P_{\tau}$,
starting at the root node of $\tau$, such that stolen subtasks of $\tau$
correspond to off-path right children of $P_{\tau}$.
Each node $v$ of $P_{\tau}$ has a collection of $O(1)$ local variables that are stored
contiguously on $S_{\tau}$; we refer to the locations taken by these variables
as the \emph{segment} for $v$, which we denote by $\sigma_v$.
The only segments that can be accessed by the stolen subtasks of $P_{\tau}$ are
the segments for nodes on $P_{\tau}$.
In addition, these segments occupy disjoint portions of $S_{\tau}$.
As each of the variables stored on $S_{\tau}$
is a limited access variable,
it follows that $\beta$ can be accessed $O(\min\{B, \operatorname{ht}(\tau)\})$ times by the stolen subtasks,
for, as already noted, $\tau$ uses $O(\operatorname{ht}(\tau))$ space.

Each time a stolen subtask accesses $\beta$ there may be a need to transfer $\beta$; further, following
this, there may be a need to transfer $\beta$ back to the processor executing $\tau$.
This causes $O(\min\{B, \operatorname{ht}(\tau)\})$ transfers of $\beta$.
In addition, if the execution of $\tau$ shifts from one processor to another, this may entail
further transfers of $\beta$.
We call such a transfer a \emph{usurpation}. It can occur at a join, if the processor $C'$
executing a stolen subtask ending at this join is the last of the two joining tasks to finish;
then $C'$ continues the work on $\tau$.
But usurpations can only occur on the up-pass of the computation, and at this point
the only further change to $S_{\tau}$ is to shrink.
As the variables on $S_{\tau}$ are limited access,
this means that there can be only  $O(\min\{B, \operatorname{ht}(\tau)\})$ usurpations
that involve accesses to $\beta$ and hence further transfers of $\beta$,
namely one for each join with a stolen subtask of $\tau$.
This is $O(\min\{B, \operatorname{ht}(\tau)\})$ transfers in total.
\end{proof}

\begin{remark}
\label{rem:padded-BP}
In our companion paper~\cite{CR10a}, we introduce \emph{padded BP} algorithms.
They are a variant of BP algorithms (specified in
Section~\ref{sec:sharper-anal}),
which in turn are a variant of
limited access Tree Algorithms.
In padded BP algorithms, each node $v$ declares an array of size
$\sqrt{r}$, where $r$ is the size of the subtask which starts at node $v$.
These arrays are otherwise unused; their purpose is to reduce the number
of block misses.
Then the bound in the above lemma changes to $O(\min\{B, \sqrt{|\tau|}\})$
\emph{(}for in padded BP algorithms, the sizes of the nodes are geometrically decreasing
as one descends the tree\emph{)}.
\end{remark}

\noindent{\bf Notation}.
Let $v$ be the node in $D$ initiating task $\tau$, where $\tau$ is either the original
task or a stolen task.
Sometimes, we write $v_{\tau}$ for $v$.
We use both
$\sigma_v$ and $\sigma_{\tau}$ to designate the segment on $S_{\tau}$ for node $v_{\tau}$.

\begin{lemma}
\label{lem:limited-write-intf-first}
Let $\cal A$ be a limited-access, top-dominant, Type 2
Hierarchical Tree Algorithm.
Let $\tau$ be either the original task in the computation of
$\cal A$ or a task which is stolen during the
execution of $\cal A$.
Let $\beta$
be a block used for
$\tau$'s execution stack $S_{\tau}$.
Then the number of transfers of block $\beta$
during the execution of $\tau$
is bounded by
\begin{eqnarray*}
Y(|\tau|,B) = \left\{
\begin{array}{ll}
O(cB) & \text{if}~{S^l}[s(|\tau|]) \ge B \\
O( \sum_{i \ge 0} c^{i}\cdot {S^l}(s^{(i)}[|\tau|)] ) & \text{otherwise}
\end{array}
\right.
\end{eqnarray*}
where $\Theta({S^l}(x))$ is a tight bound on the space used by a recursive task of size $x$
for its local variables in algorithm $\cal A$.

If $S^l(n) = \Theta(n)$, the bound becomes
\begin{eqnarray*}
Y(|\tau|, B) = \left\{
\begin{array}{ll}
O(cB) & \text{if}~s(|\tau|) \ge B \\
 O( \sum_{i \ge 0}  c^{i}\cdot s^{(i)}(|\tau|)) & \text{otherwise}
\end{array}
\right.
\end{eqnarray*}
If $s(n)\le (1-\gamma)n/c$ this is an $O(\min\{cB, |\tau|\})$ bound.
\end{lemma}
\begin{proof}
As in the proof of Lemma \ref{lem:BP-block-miss},
by Lemma \ref{lem:io-block-miss-bdd},
it suffices to bound the number of accesses by
stolen subtasks of $\tau$ to block $\beta$
plus the number of usurpations.

Again, there is a single path $P_{\tau}$ in the computation dag $D$,
such that the only segment accessed by a stolen subtask $\tau'$ of $\tau$
is the segment on $S_{\tau}$ corresponding to
the parent of $v_{\tau'}$ on $P_{\tau}$.
However, as the segments
for successive nodes on $P_{\tau}$ may reuse space on $S_{\tau}$
(if they are present for disjoint time intervals),  conceivably
the sum of the lengths, and hence
the number of
accesses to, the portions of these segments
in $\beta$ is $\Omega(B)$.
It remains to bound the sum of these lengths.

Of the blocks storing portions of  $S_{\tau}$,
we focus on those
for which the variables it stores (as opposed to their values)
may change over the course of $\tau$'s execution.
These will be blocks holding segments (or portions of segments)
whose lifetime is shorter than
$\tau$'s, i.e.~they are for nodes that are strict descendants of $v_{\tau}$ on $P_{\tau}$.
For any other block $\beta$ storing portions of $\sigma_{\tau}$,
there can be only $O(\min\{B, S^l(|\tau|)\})$ accesses to $\beta$
during $\tau$'s execution, by the limited access property.

Next, we explain the sequence of segments for Type 2 tasks that can be present
simultaneously on $S_{\tau} \cap \beta$ and that correspond to nodes on $P_{\tau}$
(we call these Type 2  segments henceforth).
There is the segment $\sigma_{\tau}$ for $\tau$, followed by
a segment $\sigma_{\tau_1}$  for $\tau_1$, where ${\tau_1}$  is called recursively
by $\tau$, followed by
a segment $\sigma_{\tau_2}$  for $\tau_2$, where ${\tau_2}$  is called recursively
by $\tau_1$, and so forth.
If $c>1$, over time there will be up to $c$ distinct $\tau_1$, one from each collection of recursive
calls, up to $c^2$ distinct $\tau_2$, and so forth.
These $c$ $\tau_1$ will reuse the same space on $S_{\tau}$, as will the $c^2$ $\tau_2$, etc.
However, segments present simultaneously use disjoint space.

Thus,
if the segment for each $\tau_1$ uses at least $B$ space, then
$\beta$ is overlapped by just $\sigma_{\tau}$ and the $\sigma_{\tau_1}$.
In this case,
the sum of the space used in $\beta$ by each of these Type 2 segments is $O(cB)$.
Otherwise, during such an interval of time, the sum of the
space used in $\beta$ by the Type 2 segments is bounded by:
$O(  \sum_{i\ge 0} c^i \cdot S^l[ s^{(i)}(|\tau|)])$.

We still need to account for the space used by tree tasks whose segments are on
$S_{\tau} \cap \beta$. We will bound this by $O(\min\{B, S^l(|\tau|)\})$
plus a constant times the space used by the Type 2 segments.
To this end, we charge the accesses in $\beta$ to the segment for a tree task $\nu$
to the segment $\sigma_{\mu}$ for the Type 2 task $\mu$ that called $\nu$.
By Property~\ref{prop:space-use}, this charge is $O(S^l(\sigma_{\mu}))$
(this is where top dominance,
the lower bound on $S^l(|\mu|)$ is used).
Since we are concerned with accesses to block $\beta$, this charging is legitimate
only if $\sigma_{\mu}$ lies fully in $\beta$.
This need not be the case for one segment, namely the segment, if any,
which overlaps $\beta$
and its predecessor block $\beta'$ on $S_{\tau}$;
call this segment $\sigma_{\overline{\mu}}$.
Instead of charging
$\sigma_{\overline{\mu}}$,
the charges for the $O(1)$ tree tasks
called by
$\sigma_{\overline{\mu}}$
are paid for directly.
The argument is the same as in Lemma \ref{lem:BP-block-miss};
each tree task $\nu$ called by
$\sigma_{\overline{\mu}}$
uses space
$O(\min\{B, \operatorname{ht}(\nu|)\})$ on $\beta$;
as $\cal{A}$ is top-dominant this is
$O(S^l(\sigma_{\overline{\mu}})) = O(\min\{B, S^l(|\tau|)\})$.
The remaining charged segments are fully contained in block $\beta$,
and so total cost for accesses to tree segments in $\beta$ is $O(\min\{B, S^l(|\tau|)\})$
plus (a constant times) the space used by the Type 2 segments.

Again, we need to account for the effect of usurpations.
But as in the proof of Lemma~\ref{lem:BP-block-miss},
the number of usurpations that cause transfers of block $\beta$ is bounded by the
number of variables stored on $\beta$ during the execution of $\tau$
and we have already bounded this quantity.
So this does not affect the overall bound.
\end{proof}
This result could be extended to higher type recursive algorithms.
To prove it for type $i+1$ algorithms, the tree computations are replaced
by type $i$ computations in the above argument, which is otherwise unchanged.

\begin{remark}
\label{rem:padded-hbp}
The bound in the above lemma applies even if the tree algorithms are padded BP algorithms,
for $S^p(|\nu|) =O(\sqrt{\nu})$ for a padded BP tree algorithm task $\nu$, and this
will be upper bounded by $S^l(|\mu|)$ for the Type 2 task $\mu$ calling $\nu$.
\end{remark}

In all our Type 2 algorithms, $ S^l(n), S^p(n)=\Theta(n)$.

\begin{remark}
When employing an algorithm $\cal A$ as a subroutine,
where $\cal A$ reads its input repeatedly,
assuming $\cal A$'s input was generated by the calling procedure, we need to ensure
$\cal A$'s input is only read $O(1)$ times, at least if the call to $\cal A$ can run
in parallel to other work.
% RJC This also arises for the depth log^2 n MM algorithm.
%
Our Matrix Multiply algorithms are examples
for which this arises.
One way of ensuring this for recursive tasks is for them
to use local variables to copy their input.
\end{remark}

\subsection{Algorithm Examples}
\label{sec:alg-ex-bl-miss}

For the MM algorithms, $S^l(n^2) = \Theta(n^2)$ (this is the space used to store the results
of the MM subproblem being computed by the task.)

The MM algorithms have the feature that
for matrices in the BI format,
each stolen subtask writes to $O(1)$ blocks
shared with its parent task, and consequently by Lemma~\ref{lem:limited-write-intf-first}
induces an additional $O(B)$ delay, measured in cache miss units.
The following lemma is immediate.

\begin{lemma}
\label{lem:MM-block-miss-cost}
The MM algorithms incur delay $O(S\cdot B)$ due to the block misses if they undergo $S$ steals.
\end{lemma}

Indeed, this is a design principle that is followed in the already mentioned
oblivious sorting algorithm~\cite{CR10b}
 and in the algorithms described in our companion paper~\cite{CR10a}: ensure
that each subtask accesses only $O(1)$ writable blocks shared with other tasks.
(One could generalize this to any bound $Z$ with a proportionate increase
in the overall costs
of $O(S \cdot Z \cdot B)$
due to the block misses, of course).

We can now explain our algorithms for converting between the RM and BI formats for storing
matrices.

To go from RM to BI,
the straightforward algorithm which recursively copies each quadrant using a tree computation
suffices.
it results in $T_{\infty}=O(\log n)$, $W=O(n^2)$, $Q=n^2/B$.

Let $\tau_{\kappa}$ denote the kernel of task $\tau$ (the remainder of $\tau$ when
stolen subtasks are removed).
The original task $\tau$ or a stolen task $\tau$ each
have a cache miss cost of $O(|\tau_{\kappa}|/B + \sqrt{\tau} + 1)$,
as they are reading from a size $\sqrt{\tau} \times \sqrt{\tau}$ submatrix in RM format.
Summed over all stolen tasks, plus the original task, this is
$O(n^2/B + \sum_{\tau \mbox{ stolen}} \sqrt{\tau})$.
The additional block delay caused by $\tau$ being stolen is $O(B)$ as $\tau$ is
writing in left to right order into a vector, the BI format for the matrix, and
so the only blocks on which it has access conflicts are the leftmost and rightmost
blocks to which it writes.

\begin{lemma}
\label{lem:BI-to-RM-compl}
The above algorithm for converting RM to BI format incurs
$O(n^2/B + n\sqrt{S})$ cache misses and a delay of $O(S\cdot B)$ due to its block misses.
\end{lemma}
\begin{proof}
The number of cache misses is maximized if the sizes of the stolen tasks are
as large as possible, namely of size $\Theta(n^2/S)$ or larger.
This causes $\sum_{\tau \mbox{ stolen}} \sqrt{|\tau}|) = O(n\sqrt{S})$.

The bound on the block delay is immediate.
\end{proof}
\noindent
{\bf Comment}.
We observe that $n^2/B + n\sqrt{S} = O(n^2/B + S\cdot B)$.

Consequently, the
cache miss cost is bounded by the sequential cache miss cost plus the block miss delay.
\vone

However, going from BI to RM is not a symmetric process; the direct logarithmic depth
tree algorithm incurs many block misses because a subtask $\tau$ may write to
$\Theta(\sqrt{|\tau|})$ blocks shared with other tasks.
But this computation is being used
for matrix multiply, so we can afford an algorithm with $T_{\infty}=O(\log^2 n)$.
This slower runtime allows us to sharply reduce the number of block misses.

This algorithm divides the length $n^2$ BI representation array into four parts,
each of which it recursively converts to RM order.
Then, using a tree computation, it copies the four subarrays into one subarray in RM order.
For this algorithm, $W=O(n^2\log n)$ and $Q=O(\frac{n^2}{B}\frac{\log n^2}{\log M})$.

In this tree computation, each task reads from two arrays in RM order to produce its output
in one array again in RM order.

\begin{lemma}
\label{lem:c-miss-BI-RM}
The above algorithm for converting BI to RM format incurs
$O(\frac{n^2}{B} S)$ cache misses and a delay of $O(S\cdot B)$ due to its block misses.
\end{lemma}
\begin{proof}
For each stolen task $\tau$ with kernel $\tau_{\kappa}$,
there are $O(\ceil{|\tau_{\kappa}|/B})$ cache misses.
The $k$ stolen subtasks at a given level of recursion, of total size $r\le n^2$, incur
$O(k+ r/B)$ cache misses. The number of cache misses is maximized, if at each level,
starting at the highest level,
each stolen task is as large as possible and between them they have combined size $n^2$.
This yields $O(\frac{n^2}{B}\log S)$ cache misses.

Again, the cost of the block misses is $O(B)$ per steal.
\end{proof}

An  improved
method for BI to RM conversion with $T_{\infty}= O(\log n)$
is given in \cite{CR10a}.

As we will see, the number of steals in the MM algorithms dominate those for
the above BI to RM
algorithm, and consequently the MM algorithms dominate the BI to RM algorithm
in terms of operation count, runtime, and combined cache miss and block delay.
Thus the just described conversion algorithms can be used with the
MM algorithms without
affecting the asymptotic complexity of the MM algorithms.

\section{The Analysis of RWS with False Sharing}
\label{sec:rand-work-steal}

Here we analyze
the performance of randomized work-stealing \cite{ABB02,BL99}
when the cost of block misses is incorporated.
Our analysis follows the approach taken in~\cite{ABB02},
but
without the assumptions, made in~\cite{ABB02},
of an $O(1)$ block size and no false-sharing.
We note that even if $B=O(1)$, when false sharing is allowed,
showing an $O(B)$ ($=O(1)$)
bound on the block miss delay
in serving
any access request to a block appears to require a non-trivial justification,
as presented in Section~\ref{sec:lim-acc}.
\hide{
Also, while it lies beyond the scope of the current paper,
we note that an $O(B)$ block miss bound could be achieved
by requiring that accesses to each block be queued.
} % end hide

We assume that a successful steal takes between $s$ and $a_2 s$ time,
for some constant $a_2 \ge 1$,
and an unsuccessful one takes
$O(s)$ time (prior work assumed both took $\Theta(s)$ time).
We assume that $s \ge b$, which seems plausible
for each steal requires reading data on another processor
and consequently a cache miss seems to be unavoidable.
Also, henceforth, for simplicity, we assume that $s$ is an integer multiple of $b$.

As in~\cite{ABB02}, we bound the number of steals by using a potential function $\phi$,
which we now define.
We assign a cost to each node in the execution dag $D$ for a given computation.
To this end,
recall that $E$ is
an upper bound, measured in cache misses, on the delay due to cache and block misses
occurring in the execution of any one node, and let
$e_1$
be an upper bound on the
number of operations (reads, writes and computations) performed in the execution of any one node.
By assumption,
$e_1=O(1)$.
Each node is given a cost of $e_1 + bE$. % $e + bD_b$.
In addition, to cover the cost of steals, any node performing a fork is
given an additional cost of $2s$ (the factor of 2 simplifies the analysis).
(This additional cost incorporates all the delay incurred by the fork 
including any block misses that may ensue.)
The cost of a path in $D$ is simply the sum of the costs of the
nodes on the path.

The height $h(u)$ of a vertex $u$ in $D$
is $1/s$ times
the maximum cost among all
the paths descending from  $u$.
$h(t)$ denotes the height of the root $t$ of $D$.
Note that
$h(t)=O(\frac 1s(e_1 + bE +s]) T_{\infty}) = O([\frac bsE + 1] T_{\infty})$,
where $T_{\infty}$ denotes the length, in vertices,
of the longest path in $D$.

We view each task corresponding to a node as performing up to
$e_1 + bE$ ``work units''
when it is executed, each work unit corresponding to one unit of time 
being expended
on its execution. 
This includes time spend waiting due to cache and block misses.

If the task $\tau_u$
associated with vertex $u$ is on a task queue, $u$
has an associated potential
$\phi(u) = 2^{1+ h(u)}$;
if $\tau_u$
is currently being executed by a processor,
with $x$ of its work units already having been performed,
$u$ has potential
$\phi(u) = 2^{h(u) - (x/s)}$;
otherwise, $u$'s potential is zero.
$\phi = \sum_u \phi(u)$.

To show progress, we analyze the algorithm in periods called \emph{phases}.
We identify two types of phases, \emph{steal} and \emph{computation} phases.
At the start of a new phase, if at least half the potential $\phi$  is
associated with vertices $u$ whose associated tasks $\tau_u$
are on task queues, this is a steal phase.
Otherwise, it is a computation phase.
A steal phase lasts until $2p$ attempted steals complete, successfully or not,
while a computation phase lasts for $b$ time units.
We show that the expected value of the potential function $\phi$
decreases at least in proportion to the number of successful steals.
\hide{
To this end, we identify two types of phases, \emph{steal} and \emph{computation} phases.
At the start of a new phase, if at least half the potential $\phi$  is
associated with vertices $u$ whose associated tasks $\tau_u$
are on task queues, this is a steal phase.
Otherwise, it is a computation phase.

A steal phase lasts until $2p$ attempted steals complete, successfully or not.
}

\begin{lemma}
\label{lem:steal-red}
In a steal phase, the expected value of $\phi$ reduces to at most $\frac78$ of its starting value.
\end{lemma}
\begin{proof}
Potential of at least $\frac{\phi}{3}$ is associated with tasks
at the heads of queues,
since, on any task queue, the heights of successive tasks decrease 
by a factor of at least
2,
and hence at least $\frac23$ of the potential associated with tasks on task queues is for tasks
at the heads of these queues.
Let $\tau_u$
be a task at the head of a task queue.
The probability that $\tau_u$
is not stolen in one attempted steal is
$1-1/p$. Hence over the at least $2p$ attempted steals, it is not stolen with
probability $(1-1/p)^{2p}\le 1/e^2$, and hence is stolen 
with probability more than
$\frac34$.
If $\tau_u$ is stolen, the potential $\phi(u)$ decreases by a factor of 2.
Consequently, the expected value of $\phi$ is reduced to at most
$\frac23 \phi + \frac14 \cdot \frac{\phi}{3} +
          \frac34\cdot \frac12 \cdot \frac{\phi}{3} = \frac{21}{24} \phi$.
\end{proof}
\begin{corollary}
\label{cor:steal-succ-prob}
With probability at least $\frac{1}{16}$, in a steal phase $\phi$ reduces to
at most $ \frac{15}{16}$ of its starting value.
\end{corollary}
\begin{proof}
Otherwise, the expected decrease is less than
$\frac{1}{16}\cdot 1 + \left( 1 - \frac{1}{16} \right) \frac{1}{16} < \frac{1}{8} $.
\end{proof}

A computation phase lasts for $b$ time units.

\begin{lemma}
\label{lem:comp-red}
In a computation phase, $\phi $ reduces to at most
$\left(1 -  \frac {b}{4s}\right)$
of its starting value.
\end{lemma}
\begin{proof}
Suppose processor $C$ is currently executing task
$\tau_u$ corresponding to vertex $u$.
Then in the current phase, $C$ can do one of three things.\\
a. It could complete its task with nothing left on its task queue.
Then the associated potential is reduced to zero.
\\
b. It could perform a fork. We show that this reduces the
associated potential $\phi(u)$ to at most $\frac34 \phi(u)$.
When a processor executing task $\tau_u$ forks,
it creates tasks $\tau_v$ and $\tau_w$,
placing $\tau_w$ on its task queue.
Recall that each forking node is assigned
an additional cost of $2s$.
Hence, the forked task $v$ that is placed on the task queue has potential
$\phi(u)/2$, and the forked task $w$ that continues the execution has
potential $\phi(u)/4$.
Thus, the potential is reduced from
$\phi(u)$ to $\phi(v)+\phi(w)\le \phi(u)/4+ \phi(u)/2 =\frac34 \phi(u)$.
\\
c. If (a) and (b) do not hold, then processor $C$ executes its task throughout
the phase without forking.
Hence processor $C$
performs a sequence of at least $b$ work units.
This reduces the starting potential $\phi(u)$ to at most
$\phi(u) 2^{-b/s} = \phi(u) (1+1)^{-b/s}  \le \phi(u)(1 -\frac{b}{2s}) $ if $b\le s$.
(Note that for $0\le x \le 1$, on setting $x'=1-x$, we have
$(1+1)^{-x} =(1+1)^{-1+x'}=\frac 12 (1+1)^{x'}\le \frac 12 [1 + x' - x'(1-x')/2! + x(1-x')(2-x')/3! + \cdots]
\le \frac 12 [ 1 + x' ] \le 1 - x/2$.)

Hence in one computation phase the potential is reduced to at most
$\frac{\phi}{2} + \left(1 -\frac{b}{2s} \right) \frac{\phi}{2} = \left(1 -\frac{b}{4s} \right)\phi$.
\end{proof}

\begin{theorem}
\label{th:succ-steal}
For $a = \omega(1)$,
with probability $\left( {1 - {2^{-\Theta(  ah(t))}}} \right)$,
the number of successful steals is bounded by
$O(p \cdot h(t)[1+a])$. %, for $h(t) \ge \log  n$.
In addition, the time spend by all the processors collectively on steals, successful and unsuccessful,
is $O(p \cdot s \cdot h(t) [1 + a])$.
\hide{
With probability $\left( {1 - {2^{ - a}}} \right)$ the number of successful steals is bounded by
$O({p({a + h(t)})})$.
In addition, the time spend by all the processors collectively on steals, successful and unsuccessful,
is $O(p [a + h(t)])$.
} % end hide
\end{theorem}
Recall that
$h(t) = O([\frac bs E + 1] T_{\infty})$.
\begin{proof}
The initial value of $\phi $ is
${2^{h(t)}}$.
While one node remains unexecuted, $\phi \ge 1$.
Thus once $\phi$ reduces to 1, the computation completes in a further 
$O(1)$ time, in
which time only $O(p)$ attempted steals can complete.
So to bound the number of attempted steals, it will suffice to consider the
time during which $\phi$ reduces from its initial value to 1.

Say that a steal phase is successful if $\phi $ reduces to at most 
$\frac{15}{16}$
of its value at the start of the phase, and that it is unsuccessful otherwise.
By Corollary \ref{cor:steal-succ-prob},
a steal phase is successful with probability at least $ \frac{1}{16}$.

Suppose that there are $x$ successful steal phases, $y$ unsuccessful ones, 
and $z$
computation phases,
until all the successful steals complete.
Then
$x + \frac{b}{{{s}}}z = O({h(t)})$.

Now each computation phase takes $b$ time units and hence uses $O({pb})$
time over all $p$ processors.
So the $z$ computation phases use $O({pbz})$ time units over all $p$ processors.
As a successful steal takes at least $s$ time units, there can be only
$O({pbz / {s}})$ successful steals
that start and finish in a contiguous sequence of computation phases.
Any other successful steal either starts or ends during a steal phase; there can be at most $2p$
of these per steal phase, $O( {p ({x + y})})$ in total.

Finally, a steal phase is unsuccessful with probability at most $\frac{15}{16}$.
By a standard computation (which asks what is the probability of fewer than
$x$ successful coin tosses in a sequence of $(a+1)x$ coin tosses),
with probability ${1 - {2^{ -\Theta( ax)}}}$, there are
$O(x\cdot a)$
unsuccessful steal phases, which yields a total of
$O\left( {\left( {x (1 + a)+ \frac{b}{{{s}}}z } \right)p} \right)$
steals with probability ${1 - {2^{  -\Theta( ax) }}}$.

To bound the time spend on steals we note that, summed over all the processors,
a steal phase uses time $O(sp)$ for the steals and a computation phase 
$O(bp)$ time.
Thus the time cost of the steals is 
$O((x+y)ps + zpb) = O((x(1 + a) +\frac{b}{s}z )ps)= O(h(t)(1+a)ps)$.
\hide{
Finally, a steal phase is unsuccessful with probability at most $\frac{15}{16}$.
By a Chernoff bound, with probability ${1 - {2^{ - a}}}$, there are $O({x + a})$
unsuccessful steal phases, which yields a total of
$O\left( {\left( {x + \frac{b}{{{s}}}z + a} \right)p} \right)$
successful steals with probability ${1 - {2^{ - a}}}$.

To bound the time spend on steals we note that, summed over all the processors,
a steal phase uses time $O(sp)$ for the steals and a computation phase $O(bp)$ time.
Thus the time cost of the steals is $O((x+y)ps + zpb) = O((x +\frac{b}{s}z +a)ps)= O(h(t)+a)ps)$.
} % end hide
\end{proof}

\hide{
Next, we address the bounds on running time that can be shown using the above result.
Recall the notation specified in Section~\ref{sec:results}:
Let $\cal A$ be an algorithm with a series-parallel computation dag $D$
being executed using RWS.
Let $T_{\infty}$ denote the length in nodes of the longest path in $D$.
Recall that
$e_1 =O(1)$
is the number of operations performed at a single node
and $E$ is a bound on the cost, measured in cache miss units, of executing a single node in $D$.
Recall  that $h(t) =  O([ \frac bs E + 1]T_{\infty})$.
Suppose that in a sequential execution, in the worst case, 
$\cal A$ performs $W$ operations
and incurs $Q$ cache misses.
Next, let $C(M,B)$ denote an upper bound on the ``excess'' cache misses 
incurred by any stolen task,
the number of cache misses over and above the number of misses 
incurred by this task
in a sequential execution of $\cal A$.
Let $Z(B)$ denote an upper bound on the delay due to block misses 
incurred by any stolen task,
measured in cache misses.
Let $D_b$ be a bound on the cost, measured in
cache misses, incurred
due to block misses
on any single path in any execution of $D$.
Finally, recall that $E$ was the bound on the cost of block misses at a single node.
} % end hide

\hide{
\begin{theorem}
\label{thm:basic-rws-bound}
When scheduled under RWS, with probability $1 -2^{-a}$, $\cal A$ runs in time
\[
O\left( \frac{W}{p} + b\frac{Q}{p}
             + \left[T_{\infty} +\frac {b}{s}E T_{\infty}    + a \right]
                                \left[ s + bC(M,B) + bZ(B) \right]
 \right).
\]
\end{theorem}
This bound, modulo changes in parameter names, assuming a zero cost 
for block misses,
becomes the bound obtained in~\cite{ABB02}.
} % end hide

In all the algorithms we consider, 
%$C(M,B) =\frac MB + \sqrt{M}$, $Z(B) = O(B)$, and
$E = O(B)$.
However, this bound need not hold in general.
\hide{
Assuming no false sharing and $B=O(1)$, Frigo and Strumpen~\cite{FS06} obtain similar bounds.
(They require that the costs in cache misses of stolen subtasks are a concave function of the size
of the subtask. We preferred to explicitly measure ``excess'' costs, but the effect is similar.)
}

\section{Analysis of HBP Algorithms}
\label{sec:sharper-anal}

HBP algorithms are obtained by imposing a further modest restriction on the Hierarchical Tree algorithms.
First, we define Balanced Parallel (BP) algorithms.
These are Tree Algorithms, but with the further restriction that they
have roughly
equal-sized subproblems; for a tree computation $T$, where $\tau$ is the corresponding
task, there are constants
$c_1 \le 1 \le c_2$ and $\alpha < 1$,
such that subtasks corresponding to the subtrees at the $i$th level all have
sizes between $c_1|\tau| \alpha^i$ and $c_2|\tau| \alpha^i$.
Next, we define Type $i$ Hierarchical Balanced Parallel (HBP) Algorithms, which correspond to
Type $i$ Hierarchical Tree Algorithms.
Here, we apply a similar restriction
to the trees forking recursive computations,
namely that the number of leaves in the subtrees at a given level are all
within a constant factor of each other
(\cite{CR10a} also requires the size of the recursive subproblems to be within a constant
factor of each other, but this is not needed for the analysis below).

As it suffices to analyze BP and Type 2 HBP algorithms to also handle
the analyses in our companion
paper, we will limit the analysis below to these classes.
(As it happens, this analysis allows us to handle the Type 3
algorithm for list ranking and the Type 4 algorithm for connected components.)

Broad-brush, our analysis is similar to that in
Section~\ref{sec:rand-work-steal}.
Again, we associate a ``height'' or \emph{level} $h(u)$ with each node
in the computation dag $D$,
and show that the same resulting potential function decreases as before.
However, we improve the bounds from Section~\ref{sec:rand-work-steal}
by reducing the over-counting of delays due to block misses.
For example, if two processors are
competing to access a block, only one of them will be delayed
on their first access.
Our current analysis assumes both are delayed.
As it turns out, this has a substantial effect on our bounds.

We do this by introducing levels for nodes in $D$ that change
dynamically as the algorithm execution proceeds.
Each node $u$ in the dag $D$ will have a current level, $h(u)$.
$h(u)$ may decrease as $u$ is executed; it may even change for a node
that has not yet been expanded in the dynamic DAG.
The challenge in designing the level $h$ is that
for an edge $(u,v)$,
the difference
$h(u) - h(v)$ needs to be at least the time taken to perform the operations
at node $u$, which could include the effect of a block miss delay.
However, we want this difference to be large only if there really
is a block delay.
Since the nodes at which block delays occur depend on the
order of execution,
our analysis needs to account for every possible order.
All of these present challenges in setting up the $h$ function.
In our approach,
we enable sufficiently large differences when needed by dynamically
reducing the value of $h(w)$ for
a suitable subset of the
nodes that could access block $\beta$ when $\beta$ incurs a block delay;
we also reduce the $h$ value for selected descendants of such nodes $w$.

The effect is to reduce the prior bound of $h(u) = O(B \log n)$ for
a size $n$ BP computation
to $O(B+ \log n)$; in fact, the tighter bound
$h(u)=O(\log n+ \min\{n,B\})$ holds.

We begin by analyzing BP computations.
The extension to HBP computations is fairly straightforward.

\subsection{BP Algorithm Analysis}

There are two classes of delays we need to consider.
The first are due to accesses to global arrays;
by Lemma ~\ref{lem:limited-write-intf-first},
there are $O(B)$ delays per accessed block.
BP algorithms have the following feature which
allow them to avoid a cascading series of delays along a path in
the computation dag $D$:

\begin{quote}
\label{ass:BP-cnstrt-first}
{\bf Regular Pattern for BP Global Variable Access}.
All writes to global variables (typically arrays of size $n$)
 are performed either at the leaf nodes or in the following
\emph{regular} pattern:
the $i$th node in the down-pass tree in inorder writes locations
$[a(i - 1)+1 \cdot\cdot~ a \cdot i]$ for some constant $a \ge 1$, and similarly for nodes in
the up-pass tree.
\end{quote}
Prefix-sums can be implemented as a sequence of two  BP computations with a regular pattern.

Together with the balanced subtree requirement, this feature ensures that
the only conflicts in accessing global arrays
occur at nodes in the bottom $a_1\log B$ levels of the up-pass tree,
for a suitable constant $a_1 \ge 1$,
and the same is true for nodes in the down-pass tree.
We call the
height $a_1\log B$ subtrees formed by these nodes \emph{conflict subtrees}.
The key property is that between them, the nodes in each pair of complementary conflict subtrees
(one in the down-pass tree, one in the up-pass tree)
access the same $O(1)$ blocks, and thus the cumulative delay due to accesses to
the global arrays is $O(B)$ (measured in units of cache misses).

The second class of accesses are to variables stored on the execution stack
(including hidden variables such as those for reporting the completion of a subtask).
We limit ourselves to BP algorithms which have the following additional feature
(stemming from a natural scoping of variables and the use of return value
variables); this is used in bounding the cost of these accesses.
\begin{quote}
{\bf BP Local Variable  Access}.
Writes to local variables by the task for a node $v$ are to $v$'s local variables,
and in the up-pass possibly to $u$'s local variables also, where $u$ is $v$'s parent  (the
node following $v$ in the up-pass computation).
\end{quote}

Recall that we let $e_1 =O(1)$ be a bound on the number of reads and writes
performed at any one node in $D$.
We also let $e_2$ be the number of times a writable
variable can be accessed during the computation.
Recall that, by the limited access property, $e_2 = O(1)$ also.
Finally, we let $e =\max\{e_1,e_2\}$.

We also assume that the blocks storing the execution stack for $C$ and those storing $C$'s task queue are disjoint.

Next, we define the current levels for the nodes in $D$.
We specify
four levels, $\ell_1$, $\ell_2$, $\ell_3$, $\ell_4$,
where $\ell_1$ for the effects of steals,
$\ell_2$ for the effects of global variables,
$\ell_3$ accounts for the effects of local variables,
and
$\ell_4$ for the interactions between local and global variables.
The current level, $h(u)$ of a node $u$ is given by
$h(u)= \ell_1(u) + \frac{b}{s} [\ell_2(u) + \ell_3(u) + \ell_4(u)] $.

A key property that we enforce on
the levels is that for each edge $(u,v)$ in $D$, $\ell_i(u) \ge \ell_i(v)$,
for $i=2,3,4$, and $\ell_1(u) \ge \ell_1(v)+2$.
Also, $\ell_i(u) \ge 0$ for $1 \le i \le 4$.
These properties will allow
us to show essentially the same reduction
in potential
during a computation phase as was shown previously in Lemma~\ref{lem:comp-red}
(see Lemma~\ref{lem:comp-red-ext} below).

$\ell_1 (u)$ is the simplest so we define it first.
Let $\operatorname{ht} (u)$ be the actual height of $u$ in the dag $D$, measured in edges.
Then
$\ell_1(u)=2 \operatorname{ht} (u) \ge 0$.
$\ell_1(u)$ remains unchanged throughout the computation.
Note that if $(u,v)$ is an edge in $D$, then $\ell_1(u) \ge \ell_1(v) + 2$.

\paragraph{$\ell_2$ definition and analysis}
To specify $\ell_2$ we need to define the following height $\Theta(\log B)$ \emph{conflict} subtrees $T$
at the bottom (leaf level end) of the up-pass tree.
Let depth $d+1$ be the greatest depth such that all the subtrees at depth $d+1$ have
$B-1$ or more nodes.
Then the nodes at depth $d$ are the roots of the conflict subtrees.
We define analogous conflict subtrees in the down-pass tree.
We pair complementary conflict subtrees, namely those comprising paired
fork-join nodes.
By the BP Global Variable Access property,
any access by the root of a conflict subtree $T$
to a global array can have a conflict only with other nodes in $T$ and its paired tree $T'$,
and thus in fact there are no conflicts as a result of these accesses:
for in the case of the up-pass tree $T$, the nodes below its root
will have completed before the root starts its computation, and in the case of the down-pass tree $T'$,
the nodes below the root will start their computation only after the root completes.
This conflict free property also applies to nodes nearer the root of the up-pass and down-pass trees.

\begin{lemma}
\label{lem:confl-tree-size}
A conflict subtree has at most $4\frac{c_2}{c_1}(B - 2) + 3$ nodes.
\end{lemma}
\begin{proof}
Necessarily, there is a subtree with root at depth $d+2$ with $B-2$ or fewer nodes.
Thus the largest subtree at this depth has at most $\frac{c_2}{c_1}(B-2)$ nodes.
Hence, every conflict tree has at most $4\frac{c_2}{c_1}(B - 2) + 3$ nodes.
\end{proof}

Let $\beta$ be a block used to store parts of the global array(s).
For each $\beta$ and each conflict subtree $T$ some of whose nodes access $\beta$,
every node $v$ in $T$
and in its paired subtree $T'$
has $eB$ added to the initial value of $\ell_2(v)$ (from a starting value of 0).
For short, we say that $T$ can touch $\beta$.

Whenever there is an access to block $\beta$,
for each conflict subtree $T$ with a node that can write to  $\beta$,
$\ell_2(v)$ is decremented
by one
for every node $v$ in $T$ and its paired tree $T'$.
All decrements will be by 1, so will not mention this henceforth.
Note that while the access is by some node $u$ in some conflict tree $T$,
decrements may also occur to $\ell_2(u')$ for the nodes $u'$ in
a neighboring conflict subtree $T''$.
The net effect is that for each conflict subtree $T$, $\ell_2(v)$ is identical for every node $v$ in $T$
and the paired $T'$.

The remaining nodes in the up-pass tree, those not in conflict subtrees,
are given an $\ell_2$ value of 0.

\begin{lemma}
\label{lem:bound-h2}
For $v$ in the up-pass tree, initially $\ell_2 (v) \le 4\frac{c_2}{c_1}e^2 B$,
and always $\ell_2 ( v ) \ge 0$.
\end{lemma}
\begin{proof}
A global array to which nodes write $e'$
items has at most $\lceil \frac{e'}{B} \cdot [4\frac{c_2}{c_1}(B - 2) + 3] \rceil \le 4e'\frac{c_2}{c_1}$
blocks that can be touched by conflict subtree $T$.
Recall that each node performs at most $e$ accesses.
Summing the $4e'\frac{c_2}{c_1}$ bound over all $e'$,
one per global array, yields a bound of $4e\frac{c_2}{c_1}$
blocks (since the sum of the $e'$
is at most $e$).
Each such block contributes $eB$
to the initial value of $\ell_2(v)$
for $v$ in $T$,
yielding the initial bound of $4e^2 e\frac{c_2}{c_1} B$
on $\ell_2(v)$.

To obtain the second bound, we note that for each block
$\beta $ that $T$ can touch,
by the limited access property,
there are at most $eB$
decrements to $\ell_2(v)$ for $v$ in $T$.
As there was an initial contribution of $eB$ to
$\ell_2(v)$ for $\beta$, it follows that $\ell_2(v)$
remains non-negative forever.
\end{proof}

In the down-pass tree, for each node $u$
outside of any conflict subtree,
$\ell_2(u)= 4e^2\frac{c_2}{c_1}B$.
It follows that on any path in the dag $D$, $\ell_2$ is non-increasing.

We have shown:
\begin{lemma}
\label{lem:ell-2-props}
For all nodes $v$ in the dag $D$, initially
$\ell_2 (v) \le 4 \frac{c_2}{c_1}e^2 B$,
and always $\ell_2 ( v ) \ge 0$.
Further, for every edge $(u,v)$ in $D$, $\ell_2(u) \ge \ell_2(v)$.
\end{lemma}

\paragraph{$\ell_3$ definition and analysis}
Recall that $\ell_3$ is intended to handle accesses to local variables
and these are restricted
as specified in the BP Local Variable Access Property.

We will consider a task $\tau$, which is either the initial task in
the computation or a stolen subtask.
We will be concerned with analyzing the delays in accessing local variables on
$S_{\tau}$, $\tau$'s execution stack.
Recall that the kernel of $\tau$ is the portion of $\tau$ that does not
get stolen.
Also recall the definition of a segment $\sigma_v$ for a node $v$ in the computation dag $D$:
it stores the variables declared by $v$, if $v$ is a fork or leaf node.
If $v'$ is a join node, its segment is the segment created by the corresponding fork node $v$.
The segments on $S_{\tau}$ correspond to the sequence of fork nodes in the kernel
whose corresponding join tasks have not yet been completed, plus possibly one segment at the top of the
stack for a leaf node, if this is the node in $\tau$'s kernel being executed currently.

We focus on the path $P_{\tau}$ in $D$.
Recall that this is the path of nodes starting at the root node (i.e.~the initial node)
for $\tau$'s computation,
and for which the off-path right children
correspond to the roots of $\tau$'s stolen subtasks.
This is a path in the down-pass tree.
We define a corresponding path $P'_{\tau}$ in the up-pass tree:
if $v$ is a node on $P_{\tau}$, necessarily a fork node,
then the corresponding join node $v'$ is on $P'_{\tau}$.
It is convenient to extent $P_{\tau}$ (and $P'_{\tau}$) so as to
connect these two paths, as follows.
Let $w$ be the root node for the last stolen subtask of $\tau$,
and let $v$ be $w$'s sibling.
Then $v$ plus the path of right children descending from $v$ to a leaf of the
down-pass tree forms the extension of $P_{\tau}$,  and the corresponding nodes in the up-pass
tree form the extension of $P'_{\tau}$.
This leaf node is common to both paths.

\begin{figure}[htbp]
% \begin{center}

% \input{fig1.latex}
\input{fig1.pstex_t}

\caption{\label{fig-l3}Notation for Down Pass and Up Pass Trees.}

% \end{center}
\end{figure}

We still need to identify the nodes on $P_{\tau}$ and $P'_{\tau}$ and their offpath
children more precisely.
See Figure~\ref{fig-l3}.
Let $\langle v_1, v_2, \cdots, v_k \rangle$,  be the path of nodes
forming $P_{\tau}$, descending
the down-pass tree, and including a leaf.
Let $\langle v_{r_1}, v_{r_2}, \cdots, v_{r_t} \rangle$
be the subsequence of these nodes at which a steal occur;
i.e.~their right children are the root nodes for stolen subtasks of $\tau$, and
let $\langle w_{r_1 +1}, w_{r_2 +1}, \cdots, w_{r_t +1} \rangle$
be this sequence of right children.
Let $v'_i$ be the node on $P'_{\tau}$ corresponding to $v_i$;
i.e.~$P'_{\tau}$ comprises the path $\langle v'_k, v'_{k-1}, \cdots, v'_1 \rangle$ of nodes
ascending the up-pass tree.
Let $w'_{r_s+1}$ be the off-path child of $v'_{r_s}$ in the up-pass tree, $1 \le s \le t$;
note that $w'_{r_s+1}$ corresponds to node $w_{r_s+1}$ in the down-pass tree.
Sometimes we will call $w_{r_s+1}$ the \emph{non-kernel} child of $v_{r_s}$,
and similarly for $w'_{r_s+1}$ w.r.t.~$v'_{r_s}$.

Let $\beta$ be a block storing a portion of $S_{\tau}$.
By the BP Local Variable Access Property, the only nodes that can access $\beta$ are
nodes in $\tau's$ kernel and nodes $w'_{r_s+1}$, $1 \le s \le t$.
We define $\ell_3$ so as to ensure a sufficient reduction in the overall potential
when such accesses occur.
As it happens some of these accesses will not cause a reduction to the $\ell_3$ values, but
as we will see in
the proof of Lemma~\ref{lem:comp-red-ext},
the potential associated with the nodes involved in such non-reducing accesses is small,
and thus much of the potential is for nodes whose potential does drop,
ensuring a sufficient overall drop in potential during an execution phase.

Now, we are ready to define $\ell_3$ precisely.

\vhalf
\noindent
{\bf Initial values of $\ell_3$}.

\vhalf

For a node $v'$ in the up-pass tree, initially,
\[
\ell_3(v') =2e \cdot
\mbox{length in vertices of the path from $v'$ to the root $t$ of the up-pass tree.}
\]
So $\ell_3(t) = 2e$ initially (recall that $t$ denotes the root of the up-pass tree).

Let $\ell_3(f)$ be the  maximum
initial value of $\ell_3$ over all leaves in the
up-pass tree
(which are also the leaves in the down-pass tree).

For non-leaf nodes $v$ in the down-pass tree,
\[
\ell_3(v) = \ell_3(f) + e \cdot \mbox{height of the down-pass tree }
+ e \cdot ( \mbox{height of $v$ in the down-pass tree} - 1).
\]

\noindent
{\bf Update specification}.
We now describe how $\ell_3$ is updated during the computation. In contrast
to the other $\ell_i$, these updates do not always pay for the cost of the
current block miss. In particular, block wait costs at some of the affected
$w'_{j}$ may not be charged under these updates, and in one case we do
not decrease $\ell_3$ at any node even though block wait costs can be
incurred at some nodes. In the proof of
Lemma \ref{lem:comp-red-ext}
we set up a charging scheme which we use to establish
that all of these
incurred costs are covered by our update mechanism for $\ell_3$.

\vhalf
The down-pass for $\tau$ is defined to end when only one leaf in $\tau$'s
kernel remains to be executed; this is the leaf on $P_{\tau}$.
During the down-pass, $\ell_3$ is updated as follows.

---
If there is a node on the task queue for the processor executing
the kernel of $\tau$,
then no decrement to the $\ell_3$ values occurs for accesses to $S_{\tau}$.

---
Otherwise
(i.e., the task queue is empty),
for each block $\beta$ storing a portion of $S_{\tau}$,
if an access to $\beta$ by $w'_{r_s +1}$ completes while other nodes
accessing $\beta$ undergo block misses, $\ell_3(v)$ is
decremented for all
non-completed, non-leaf proper descendants of $v_{r_s}$ in the down-pass tree.

---
The remaining possibility is that an access by $v_i$ to $\beta$ completes,
where $v_i$ is the topmost
unexecuted node on $P_{\tau}$; then only $\ell_3(v_i)$ is decremented.

\vhalf
By the time the down-pass of $\tau$ completes,
the only nodes
remaining to be executed in the kernel of $\tau$ are the nodes on $P'_{\tau}$.
Portions of stolen subtasks of $\tau$ may also still be under execution.

Note that in the up-pass, $v'_i$ is executed only after all $v'_j$, $j > i$,
have completed.
We now specify updates to $\ell_3$ during the up-pass of $\tau$.

---
Consider a vertex $v'_i$, for some $i$, $1 \leq i \leq k$.
When an access to $\beta$
by $v'_{i}$ completes,
$\ell_3(v'_i)$ is decremented,
as is $\ell_3(w'_i)$ if $w'_i$ is not completed and it is the non-kernel
child of $v'_{i-1}$
(i.e., the stolen sibling of $v'_i$). This occurs
if $v'_{i-1}$ is a node at which
a join with a stolen subtask of $\tau$ occurs.

---
When an access to block $\beta$ by $w'_{r_s +1}$ completes, for
some $s$,
$1 \le s \le t$,
the following decrements occur:
to $\ell_3(v'_j)$ for $j \ge  r_s +1$, and to $\ell_3(w'_{r_{s'}+1})$
for $s'>s$,
for non-completed $v'_j$ and $w'_{r_{s'}+1}$.

\begin{lemma}
\label{l2-decr-up-pass-tree}
Let $u'$ be the parent of $v'$ in the up-pass tree.
Then $\ell_3(v') \ge \ell_3(u') \ge 0$ always.
\end{lemma}
\begin{proof}
If $u'$ and $v'$ are both in $\tau$'s kernel, and $v'$ is not on $P'_{\tau}$, then
$v$ completes before the up-pass begins, i.e.~before $\ell_3(u')$ and $\ell_3(v')$
change from their initial values;
so in this case, $\ell_3(v') \ge \ell_3(u')$ and $\ell_3(v') \ge 0$ always.

It remains to consider nodes on $P'_{\tau}$ and their non-kernel children.
Suppose that $u' = v'_{i-1}$ for some $1 < i \le k$.
Note that there are at most $2e$ decrements to $\ell_3(v'_{i})$,
and if $w'_{i}$ is a non-kernel node, to $\ell_3(w'_{i})$,
for which there are not simultaneous decrements to  $\ell_3(v'_{i-1})$.
As a result, $\ell_3(v'_{i-1}) \le \ell_3(v'_{i}), \ell_3(w'_{i})$ always.

It follows that $\ell_3(u') \le \ell_3(v')$ always, for all parent-child pairs in the up-pass tree.

We turn to the second part of the claim.
Recall that $t$ denotes the root of the dag.
$\ell_3(t)$ is decremented only when $t$ succeeds in an access; i.e.~at most $e$ times.
So $\ell_3(t) \ge e > 0$ always. The property that $\ell_3(v') \ge 0$ follows
from the first part by induction on the depth of $v'$ in the up-pass tree.
\end{proof}

\begin{lemma}
\label{l2-decr-down-pass-tree}
Let $u$ be the parent of $v$ in the down-pass tree.
Then $\ell_3(u) \ge \ell_3(v) \ge 0$ always.
\end{lemma}
\begin{proof}
For a node $v$ at depth $d$ in the down-pass tree, there are at most
$e(d + 1)$
decrements of $\ell_3(v)$ (note that we define the root of the down-pass tree to have depth 0).
More specifically,
there are $e$ decrements at node $v$
and $e$ decrements due to each $w'_{r_s+1}$ for which $v_{r_s}$ is a
proper ancestor of $v$.

Thus if $u$ is the parent of leaves in the down-pass tree, then $\ell_3(u)$ undergoes at most
$e \cdot \mbox{height of the down-pass tree} $ decrements;
consequently, $\ell_3(u) \ge \ell_3(v)$, for $v$ a child of $u$.

And for a node $u$ that has a non-leaf child, $\ell_3(u)$ can undergo at most $e$ decrements that
its children do not face, hence to ensure that $\ell_3(u) \ge \ell_3(v)$, for $v$ a child of $u$, it suffices that
$\ell_3(u) \ge \ell_3(v)+ e$ initially.

For a leaf node $v$, $\ell_3(v) \ge 0$ follows by Lemma~\ref{l2-decr-up-pass-tree}.
For other nodes in the down-pass tree,
$\ell_3(v) \ge 0$ follows by induction on the height of $v$ in the down-pass tree.
\end{proof}

It follows from Lemmas~\ref{l2-decr-up-pass-tree} and \ref{l2-decr-down-pass-tree} that:
\begin{lemma}
\label{lem:bound-h1}
$\ell_3$ only decreases on descending the dag $D$.
Further, $\ell_3(u) = O(\log n)$ for any node $u$ in an $n$-leaf dag $D$,
and $\ell_3(u) \ge 0$ always.
\end{lemma}

Next, we bound the numbers of successful and unsuccessful steals
by means of a potential argument, using the same potential function as
in Section~\ref{sec:rand-work-steal}:
If $u$ is on a task queue, $u$
has potential
$\phi(u) = 2^{1+ h(u)}$;
if $\tau_u$
is currently being executed by a processor,
with $x$ of its work units already having been performed,
$u$ has potential
$\phi(u) = 2^{h(u) - (x/s)}$;
otherwise, $u$'s potential is zero.
$\phi = \sum_u \phi(u)$.
However, we redefine the notion of work units. Here, we allocate $b$ work units
for each read or write, which allows for a successful access to the relevant
variable, but
in the case of
an access delayed due to a block miss,
the potential will reduce only through changes to $h(u)$.

We define steal and computation phases as before,
except that a computation phase will now last for $2b$ steps;
this ensures that in a computation phase, a processor $C$ with enough work
will either complete $b$ steps of work,
or there is a successful access to a block $\beta$ by a processor competing
with $C$ for access.
The same proof as in Lemma~\ref{lem:steal-red} shows:

\begin{lemma}
\label{lem:steal-red-ext}
In a steal phase, the expected value of $\phi $ reduces to at most $\frac78$ of its starting value.
\end{lemma}

\begin{lemma}
\label{lem:comp-red-ext}
In a computation phase, $\phi $ reduces to at most
$\left({1 - \Theta \left( {b / {s}}\right)}\right)$
of its starting value.
\end{lemma}
\begin{proof}
For each node $u$ being executed there are two possible outcomes during
the phase.

\smallskip
\noindent
{\bf 1.}
$u$'s execution completes, ending with:

\begin{description}
\item[a.]
A fork.
\\
Suppose that this creates nodes $v$ and $w$ with $v$ being executed and $w$ on a task queue.
By the definition of ${\ell_1}$,
$\ell_1(u) \ge {\ell_1}(w) + 2$, ${\ell_1}(v) + 2$,
and hence
$h(u) \ge h(w) + 2, h(v) + 2$, and
$\phi(u) \ge 2\phi(w) ,4\phi(v)$.
Therefore, $\phi(v) + \phi(w) \le \frac{3}{4}\phi(u)$,
at least a
$({1 - \Theta( {b / {s}})})$  reduction.

\item[b.]
A join.
\\
Suppose that it activates node $v$.
Again, ${\ell_1}(u) \ge {\ell_1}(v) + 2$
and so $\phi(v) \le \frac{1}{4}\phi(u)$,
at least a  $({1 - \Theta({b / {s}})})$ reduction.

\item[c.]
$u$ simply ends (because its sibling node is not finished
and so a join cannot occur yet).
\\
Here the potential reduces from
$\phi(u)$ to $0$,
certainly at least a $({1 - \Theta({b / {s}})})$ reduction.
\end{description}

\noindent
{\bf 2.}
$u$ runs for $2b$ time units and completes
at least $b$ work units.

Then $\phi(u)$
decreases by at least
$2^{-b/s}$.

\vhalf
\noindent
{\bf 3.}
$u$ runs for $2b$ time units
and incurs a block miss
delay due to some other read or write succeeding on a block $\beta$.

If $\beta$ is storing part of a global array, then either $\ell_2(u)$ or $\ell_4(u)$
decreases, causing $\phi(u)$ to decrease by at least $2^{-b/s}$.
If $\beta$ is storing only local variables, then we need to consider what
happens to $\ell_3(u)$.
In this case, suppose that $\beta$ is storing a portion of
execution stack $S_{\tau}$.
$u$ need not be a part of $\tau$.

if $\ell_3(u)$ decreases,
then $h(u)$ decreases by $\frac bs$,
and hence $\phi(u)$ decreases by $2^{-b/s}$.

\vhalf
However, $\ell_3(u)$ may not decrease.
This leads to three more cases, which we consider below, and are the
most nontrivial part of the analysis.

\begin{description}
\item[a.]
$\tau$ is currently executing its up-pass.\\
Let $i$ be the least $j$ such that
$l_3(v'_j)$ is decremented.
Then the only nodes that may incur a block miss without a corresponding decrement of their
$\ell_3$ value are
$w'_{r_s + 1}$ for $r_s +1 < i$.
Now
$\phi(v'_i) + \sum _{r_s + 1 < i} \phi(w'_{r_s + 1}) \le \phi(v'_i)[1 + \frac 14 + \frac {1}{16} + \cdots] \le \frac 43 \phi(v'_i)$.
Thus the decrement to
$\phi(v'_i) + \sum _{r_s + 1 < i} \phi(w'_{r_s + 1})$
is by a factor of
at least $\frac 34\cdot \frac bs$.

\item[b.]
$\tau$ is currently executing its down-pass and some $\ell_3(v_i)$ is decremented.\\
The  only nodes that may incur a block miss without a corresponding decrement of their
$\ell_3$ value are
$w'_{r_s + 1}$, $r_s + 1 \le i$.
Now
$\phi(v_i) + \sum _{r_s + 1 \le i} \phi(w'_{r_s + 1}) \le \phi(v_i)[1 + \frac 14 + \frac {1}{16} + \cdots] \le \frac 43 \phi(v_i)$.
Thus the decrement to
$\phi(v_i) + \sum _{r_s + 1 \le i} \phi(w'_{r_s + 1})$
is by a factor of
at least $\frac 34\cdot \frac bs$.

\item[c.]
$\tau$ is currently executing its down-pass but no $\ell_3$ value
is decremented for nodes accessing $\beta$.\\
We will show below
that the total potential of nodes covered by this case is
$\frac 13 \phi$, where $\phi$ is the
total potential at the start of the current computation phase.
This will suffice, for

then it follows that the nodes covered by the other cases in (1) and (2)
have combined potential at least
$(\frac 12 - \frac 13) \phi = \frac 16 \phi$
at the start of the phase, and hence the overall potential reduction is
by a factor of
at least $\frac 16 \cdot \frac 34 \frac bs$.

Consider the nodes accessing $\beta$.
As there is no reduction to the $\ell_3$ values of nodes accessing $\beta$,
there is a node $v_i$ on $P_{\tau}$ which is on the task queue.
By the Local Variable Access Property,
the only nodes which can be accessing $\beta$ are a node $v$ in the kernel of $\tau$
and nodes $w'_{r_s+1}$ with $r_s+1 \le i$
(for nodes $w_{r_s'+1}$ with $r_s'+1 > i$ have not yet been initiated and so nor have
nodes $w'_{r_s'+1}$ with $r_s'+1 > i$).
Note that $v$ is either the sibling of $v_i$ or a descendant of that sibling.
Since $v_i$ is on the task queue, it follows that
$\phi(v_i) \ge 2 \phi(v)$.
Further, since $\ell_3$ reduces by 2 in each successive level of the up-tree,
we have
$\phi(v) +\sum_{r_s+1 \le i} \phi(w'_{r_s+1}) \le  \frac 43 \phi(v) = \frac 23 \phi(v_i)$.
Summing over all such $v$ yields a total of at most $\frac 13\phi$ potential
associated with nodes incurring
block misses on blocks $\beta$ with no ensuing change to their $\ell_3$ values.

\end{description}

\noindent
{\bf (4)}
%
% RJC 3/17 rewritten as discussed
There might appear to be
one more case to consider, namely when $\tau$ is a stolen
task which is computing at
the root of its up-pass tree, and it writes into the execution stack of its
parent $\tau'$. But in this case, $u$ will be a $w'$ node for 
$\tau'$,
% the parent task,
and 
this is already handled by
% its cost is incorporated within 
the analysis of parts 3a--3c above for the parent task $\tau'$.
\end{proof}

\hide{
\noindent
{\bf Comment.}  One might ask whether Case 2c in the above proof can be handled only by having
no decrement to $\ell_3$ values, or whether a different definition of $\ell_3$ values would allow
a more direct argument. Were the decrement of some $\ell_3$ values to cover this case, it would appear
that $\ell_3(v)$ would need to be decremented.
But consider the case that $v$ were the child of a node $v'_i$ on $P'_{\tau}$.
$v$ could undergo $\Theta(B)$ decrements.
To accommodate this, the initial values of $\ell_3$ would have to satisfy
$\ell_3(v) \ge \ell_3(v_i) + \Theta(B)$ (an additive term of $eB$ would suffice).
However, prior to the execution, it is not known which nodes are  on $P_{\tau}$ and
$P'_{\tau}$, and it is not clear that it is legitimate to specify them upfront given the randomized
nature of the analysis.
This is not to rule out the possibility of a construction where Case 2c is handled by decreases
in $\ell_3$ values.
}% end hide

\paragraph{$\ell_4$ definition and analysis}
%\marginpar{RJC 3/3}
% RJC some refinements are needed here.

Conceivably, there is one block storing a part of a global array
which has not yet
been accounted for. This is the block, if any, that is being shared
between a global array and the
execution stack for the task initiating the BP computation.
(While this may seem contrived in the context of a single BP computation,
this is a definite possibility when the BP computation is a subroutine
of a larger HBP
computation;
 the local variables declared in the HBP computation are global
variables from the perspective of the BP computation.)

As there is just one such block, $\beta$, we account for it by
initializing $\ell_4(u)$ to $eB$ for
every node in $D$.
Whenever
% RJC
an access to $\beta$ causes a block miss,
$\beta$ is accessed,
$\ell_4$ is decremented for every node in $D$.
% RJC new sentence
For specificity, we define the block miss due to a core $C$'s successful
writes to occur on the last write by $C$ prior to the block's transfer to
another core (we make this definition to avoid the possibility of distinct
accesses by core $C$ being deemed to cause the block misses by cores $C_1$, $C_2$, $\cdots$,
when these misses are all due to a single uninterrupted sequence of accesses
by $C$).
Thus $\ell_4(u)$ is the same for every node $u$ in $D$,
and as this decrement occurs at most $eB$ times, $\ell_4(u) \ge 0$ always.
The following lemma is immediate.

\begin{lemma}
\label{lem:ell-4-props}
For all nodes $v$ in the dag $D$, initially $\ell_4 (v) = e B$,
and always $\ell_4 ( v ) \ge 0$.
Further, for every edge $(u,v)$, $\ell_4(u) = \ell_4(v)$.
\end{lemma}

\vhalf

The next theorem follows by the same proof as for
Theorem~\ref{th:succ-steal}.
\begin{theorem}
\label{th:succ-steal-ext}
For $a = \omega(1)$,
with probability $\left( {1 - {2^{ - \Theta(a h(t)}}} \right)$, the number of successful steals is bounded by
$O(p [1+a]  h(t))$,  where $h(t)$ denotes the height of the root $t$ of $D$.
In addition, the time spend by all the processors collectively on steals, successful and unsuccessful,
is $O(p \cdot s[1+a] h(t))$.
\end{theorem}
Recall that $h(t) = \frac{b}{s}[\ell_3(t) + \ell_2(t)] + \ell_1(t)  = O(\frac{b+s}{s}\log n + \frac bs B)$,
using Lemmas~\ref{lem:bound-h2} and~\ref{lem:bound-h1}.
This contrasts with the bound for $h(t)$ of $O([1 + \frac bs] B\log n)$ used in the earlier Theorem~\ref{th:succ-steal}
(on taking $E=O(B)$ as given by Lemma~\ref{lem:limited-write-intf-first}).

\subsection{HBP computations}
\label{sec:hbp-anal}

The rules restricting the writes in BP computations apply equally to
the down-pass
and up-pass trees used to instantiate recursive calls in HBP algorithms
(see Section~\ref{sec:lim-acc} for the specification of
Type 2 (and Type $i>2$)
Hierarchical Tree Algorithms and hence of HBP algorithms).
The subgraphs corresponding to the recursive computations
are analogous to the leaves of a BP computation.
This permits us to perform an analysis of the HBP computations which is
similar to that for the BP computations.

To enable such an analysis, we require that the writes by the
recursive computations to the local
variables (arrays) of their calling procedures obey
an analog of the
Regular Pattern for BP Global Variable Access,
namely that the left-to-right sequence of recursive computations write to
successive disjoint portions of the parent's
array(s).
Further, we require that within each
recursive procedure, its
writes to these arrays be similarly constrained. A simple way of ensuring this
is to impose the following constraint:
\begin{quote}
A recursive call performs its writes to such arrays by means of a BP
computation that occurs at the end of the recursive call.

The collection of these BP computations terminating the recursive calls obeys the
\emph{Regular Pattern for Global Variable Access by a BP Collection}, namely
the $i$th node in inorder
in the down-pass tree collection writes locations
$[a(i - 1)+1 \cdot\cdot~ a \cdot i]$ for some constant $a \ge 1$, and similarly for nodes in
the up-pass tree collection.
(An inorder traversal of an ordered collection of trees traverses each tree in inorder in turn,
in the order given by the collection.)
\end{quote}
An HBP algorithm can always
be modified to have this structure, by accumulating such writes in an array local to the
recursive call which is then copied at the end of the recursive call.

We use the $\ell_i$ functions, modulo some small changes, as for the BP computations.
%\marginpar{RJC 3/10}
% RJC changing \ell_1 definition
In fact, the definition of $\ell_1$ in unchanged:
$\ell_1(u)$ is 2 times the height of $u$ in the dag $D$.
For $i\ge 2$,
in order to distinguish the definitions of $\ell_i$ for the BP algorithms and for the
HBP algorithms, henceforth we use the notation $\ell_i^{\operatorname{BP}}$ to refer to the
previously defined $\ell_i$ functions, and reserve the notation $\ell_i$
to refer to the functions used to analyze the HBP algorithms.

One way of viewing an HBP computation is to regard its computation dag as
comprising a collection of down-pass and up-pass trees joined together
either by shared leaves (in the case of the two trees forming a BP computation)
or by connecting edges (in the case of pairs of sequenced HBP computations ---
see the rules for constructing $D$ at the start of Section~\ref{sec:comp-model}).
Note that the two BP computations forming the prefix sum algorithm provide
a simple example of a sequencing.

Our goal is to use essentially the same $\ell_i$ functions as for the BP
computations for each down-pass and up-pass tree in an HBP computation, for $i\ge 2$.
But, in addition, we continue to want that $\ell_i(u) \ge \ell_i(v)$
for every edge $(u,v) \in D$.
To this end, we define the $\ell_i$ values in an HBP computation
as follows.

Let $T'$ be a down-pass tree and let $s'$ denote its root.
%\marginpar{RJC 3/10}
% RJC rewritten
For $i=2$ and 3,
% For $i=1$ and 3,
$s'$ receives $i$-increment equal
to the initial value of $\ell_i^{\operatorname{BP}}(s')$.
% RJC 3/10
% and for $i=2$, equal to the initial value of $\ell_i^{\operatorname{BP}}(s')$ plus 1.
For all other nodes, the $i$-increment is set to 0.

% RJC 3/10
For $i=2$ and 3,
for each node $x$ in the dag $D$ for the HBP computation, we define
$\Delta_i(x)$ to be the maximum, over all paths from $x$ to the bottommost vertex $t$,
of the sum of $i$-increments along the path, excluding the increment at $x$, if any.

% RJC 3/10
For $i=2$ and 3, for node $x$ in $D$, we then define
$\ell_i(x)$ to be $\ell_i^{\operatorname{BP}}(x) + \Delta_i(x)$.
$\ell_2(x)$ and $\ell_3(x)$ are updated using the same rules as for BP computations,
by decrementing $\ell_i^{\operatorname{BP}}(x)$.

%\marginpar{RJC 3/10}
% RJC I have added a proof: I am not sure if this is overkill.
\begin{lemma}
\label{lem:li-decreases-1-3}
For each edge $(u,v)$ in the dag $D$, $\ell_i(u) \ge \ell_i(v) \ge 0$ always,
for $1 \le i \le 3$.
\end{lemma}
\begin{proof}
The claim is immediate for $\ell_1$ as its value never changes.
For $\ell_2$ and $\ell_3$, the argument is identical.
We show it for $\ell_2$.

If $(u,v)$ is a tree edge, then the initial values of
$\ell_2(u)=\Delta_2(u)+\ell_2^{\operatorname{BP}}(u)=\Delta_2(v)+\ell_2^{\operatorname{BP}}(u)$.
Changes to $\ell_2(u)$ and $\ell_2(v)$ maintain the property that
$\ell_2^{\operatorname{BP}}(u) \ge \ell_2^{\operatorname{BP}}(v)$
and hence that
$\ell_2(u)= \Delta_2(v)+\ell_2^{\operatorname{BP}}(u) \ge
\Delta_2(v)+\ell_2^{\operatorname{BP}}(v) = \ell_2(v)$.

For a sequencing edge, i.e.\ a non-tree edge, 
$\Delta_2(u)$ is at least $\Delta_2(v)$
plus the initial value of $\ell_2^{\operatorname{BP}}(v)$,
for $\Delta_2(v)$ does not include node $v$'s 2-increment,
namely the initial value of $\ell_2^{\operatorname{BP}}(v)$,
while $\Delta_2(u)$ does include this 2-increment.
As $\ell_2^{\operatorname{BP}}(u)\ge 0$ always,
$\ell_2(u) \ge \Delta_2(u) \ge
\Delta_2(v) + \ell_2^{\operatorname{BP}}(v) = \ell_2(v)$ always.
\end{proof}

\paragraph{$\ell_4$ definition and analysis}
There is one substantial change in the definition of the $\ell_i$ values for HBP computations
compared to HP computations. It concerns a block $\beta$ that is shared between
the segment $\sigma_{\tau}$ and segments for descendant nodes in $D$.
The difficulty we face concerns parts of one or more arrays stored in
$\sigma_{\tau}$
%\marginpar{RJC 3/11}
% \marginpar{VLR:3/2}
%VLR: Can we call the above array global and the stored results mentioned below
%local?
%
% RJC no; but see the addition below.
which
are being used to store the results from recursive calls made by $\tau$.
% RJC
These are analogous to the global arrays in BP algorithms.
In a BP computation, block misses on $\beta$ involving
these arrays were accounted for by the $\ell^{\operatorname{BP}}_4$ level function.
% RJC
Now, by Lemma~\ref{lem:limited-write-intf-first}, the accesses to this block
could face a block delay of up to $Y(|\tau|, B)$.

We define a \emph{basic} value $\ell_4^B(v)$ for each vertex $v$ procedurally, as follows.
We start by setting the initial values of $\ell_4^B(v)$ to zero.
For each vertex $v$ in the computation dag for
a Type 2 task $\tau$, we increase the initial value of $\ell_4^B(v)$
by  $Y(|\tau|, B)$.
We say that a BP task is \emph{complete} if it is the task for the full BP computation
as opposed to a subtask.
For each vertex $v$ in the computation dag for a complete BP task $\tau$, we increase
the initial value of $\ell_4^B(v)$ by $e\min\{B, |\tau|\}$.
Note that each vertex may be part of the subdag for multiple nested tasks.
Thus the basic $\ell_4^B$ value of a vertex $v$ may be the result of
multiple positive increments, one for each Type 2 task $v$ is part of,
and one for the complete BP task $v$ is part of, if any.

We define 4-increments as follows.
Each node $v$ originating a Type 2 task or a complete BP task
receives $4$-increment equal
to $\ell_4^B(v)$; all other nodes receive a 4-increment of 0.
Next, for each node $x$ in $D$,
we define $\Delta_4(x)$ to be the maximum,
over all paths from $x$ to the bottom of $D$,
of the sum of $4$-increments along the path, excluding the increment at $x$, if any.
Finally, we define $\ell_4(x) = \ell_4^B(x) +\Delta_4(x)$.

\vhalf

Whenever an access to $\beta$ by $\tau$ or its subtasks
causes a block miss,
$\ell_4$ is decremented for every node in the subdag $D_{\tau}$
representing $\tau$'s computation.
Thus $\ell_4(u)$ is decremented by the same amount as a result of accesses
to $\beta$ for every node $u$ in $D_{\tau}$,
and this is a total decrement of at most $Y(|\tau|, B)$.

This accounts for all possible block delays.
Accordingly, we redefine $h(v) =  \ell_1(v) + \frac{b}{s}[\ell_2(v) + \ell_3(v) + \ell_4(v)]$.
%\marginpar{RJC 3/10}
% RJC changed as per my email
As $\ell_1$ is strictly decreasing on
% RJC 3/10
every edge $e=(u,v)$ in $D$, it follows that
% any edge in a tree, for a tree edge $(u,v)$,
$h(u) \ge h(v) + 2$.
\hide{
As $\ell_2(v)$ has an increment for each non-tree edge
(due to the plus 1 in the definition of 2-increments), for such edges $(u,v)$,
$h(u) \ge h(v) + \frac{b}{s}$.
} % end hide

\vhalf

We note that the definition of $\ell_4$
can be extended
to Type $j$ algorithms, $j > 2$,
by using functions % values of
$Y$ appropriate for Type $j$ algorithms.

The following result is straightforward.
\begin{lemma}
\label{lem:li-decreases}
For each edge $(u,v)$ in the dag $D$, $\ell_4(u) \ge \ell_4(v) \ge 0$ always.
\end{lemma}

The bounds for Lemmas~\ref{lem:steal-red-ext} and~\ref{lem:comp-red-ext}
extend unchanged to the HBP computations,
yielding the following Theorem.

\begin{theorem}
\label{thm:rws-work}
Let $\cal A$ be a limited-access, exactly linear space bounded, Type 2 HBP algorithm.
Also, let $h(t)$ denote the initial level of $D$'s root $t$.
Then, when scheduled under RWS,
for $a=\omega(1)$,
with probability $1 - 2^{-aT_{\infty}}$, for any integer $a \ge 1$,
$\cal A$ undergoes $O(p \cdot h(t)(1 + a))$  successful steals.
In addition, the time spend by all the processors collectively on steals,
successful and unsuccessful,
is $O(p \cdot s[1+a] h(t))$.
\end{theorem}

It remains to bound the value of the $\ell_i$ at the root of the dag $D$.
For $\ell_1$ and $\ell_3$, the values are proportional to the length of the longest
path in $D$.
The next lemma bounds the values of $\ell_2$ and $\ell_4$.

\begin{lemma}
\label{lem:hbp-bound}
Let $\cal A$ be a limited-access, top-dominant Type 2 HBP algorithm
with $S^l({\cal{A}}) = {\Theta(n)}$.
Suppose that each recursive call $\cal{A}$ makes has size at most
% RJC 3/10 reworded to be consistent with earlier lemmas
$s(n) \le n/b$, for some constant $b > 1$.
% $s(n) \le (1 - \gamma)n$, for some constant $0<\gamma<1$.
Let  $c \geq 1$ denote the number of collections of recursive calls made by $\cal A$.

Further suppose that $\cal{A}$ observes the constraints on writes for BP
and HBP computations, and let $\cal{A}$ have size $n$.
Then for every node $v$ in its computation dag $D$,
% RJC removal of ell_5 and consequent changes
\[
\ell_2(v) = O(~B\sum_{i < s^*(n,B)} c^{s^*(n,B)} + \sum_{i \ge s^*(n,B)} c^i \cdot S^l(s^{(i)}(n))~),
\]
\[
\ell_4(v) = O(~B\sum_{i < s^*(n,B)} c^{s^*(n,B)} + \sum_{i \ge s^*(n,B)} c^i \cdot Y(s^{(i)}(n),B)~),
\]
\hide{
\[
\ell_2(v), \ell_4(v) = O(~B\sum_{i < s^*(n,B)} c^{s^*(n,B)} + \sum_{i \ge s^*(n,B)} c^i \cdot S^l(s^{(i)}(n)),
\]
\[
\ell_5(v) = O(~B\sum_{i < s^*(n,B)} c^{s^*(n,B)} + \sum_{i \ge s^*(n,B)} c^i \cdot Y(s^{(i)}(n),B)~),
\]
} % end hide
where $s^*(n,B)$ is the least number $i$ of iterations of $s$ such that
$S^l(s^{(i)}(n)) < B$.
\end{lemma}
\begin{proof}
This is immediate from the recursive structure of the HBP computation.
Let $\tau$ be a task corresponding to a recursive computation in $\cal{A}$.
$\tau$ adds $O(\min\{S^l(|\tau|), B\})$ to
$\ell_2$
% RJC
% and $\ell_4$,
and  $O( Y(|\tau|, B))$ to
% RJC
$\ell_4$.
% $\ell_5$.
\end{proof}

\begin{corollary}
\label{cor:tighter-hbp-bound}
For $\cal{A}$ as specified in Lemma~\ref{lem:hbp-bound}, if $S^l(n) = \Theta(n)$,
\[
\ell_2(v),\ell_4(v) = O(~B \sum_{i < s^*(n,B)} c^{s^*(n,B)} + \sum_{i \ge s^*(n,B)}  \sum_{j\ge i} c^{j} \cdot s^{(j)}(n)~),
\]
where $s^*(n,B)$ is the least number $i$ of iterations of $s$ such that
$s^{(i)}(n) < B$.
If in addition
% RJC changes as a result of removing gamma from the lemma statement
$s(n) \le n/(bc)$, where $b > 1$ is a constant,
% $s(n) \le (1 -\gamma)n/c$, where $\gamma > 0$ is a constant,
this is an $ O(B s^*(n,B))$ bound for $c=1$ and an $O(Bc^{s^*(n,B)})$ bound for $c>1$.
\end{corollary}
\begin{proof}
By Lemma~\ref{lem:limited-write-intf-first}, for $i\ge s^*(n,B)$,
$Y(s^{(i)}(n), B) = \sum_{j\ge i} c^{j-i} s^{(j)}(n)$.
Substituting into the bound of Lemma~\ref{lem:hbp-bound} yields the result.
\end{proof}

\begin{theorem}
\label{th:path-length}
Let $\cal A$ be a limited-access, top dominant Type 2 HBP algorithm
 with $S^l(n) = \Theta(n)$.
Recall that  $c \geq 1$ denotes the number of collections of recursive calls made by $\cal A$, and that
$s(n)$ is a bound on the size of the recursive subproblems called by $\cal{A}$.
Then, $h(t)$ is bounded as follows.

(i) $c=1$:
$O(\frac{b+s}{s}T_{\infty} +  \frac bs B s^*(r,B))$, where $s^*(n,B)$ is the number of
applications of $r$ needed to reduce $n$ to at most $B$.

(ii) $c=2$ and $s(n) = \sqrt{n}$:
$O(\frac{b+s}{s}T_{\infty} +  \frac bs B\frac{\log n}{\log B})$.

(iii) $c=2$ and $s(n) = n/4$: $O(\frac{b+s}{s}T_{\infty} +  \frac bs \sqrt{nB})$.
\end{theorem}
These choices of $c$ and $s(n)$ are the ones that occur in our HBP algorithms~\cite{CR10a}.
Similar bounds are readily obtained for other values of $c$ and $r$.
%{sec:lim-acc}
\begin{proof}
(i) follows immediately from Corollary~\ref{cor:tighter-hbp-bound}.
For (ii), we have $s^*(n,B) = \log(\frac{\log n}{\log B})$; on substituting
in  the bound from Corollary~\ref{cor:tighter-hbp-bound}, the
result is immediate; similarly, for (iii),  $s^*(n,B) = \sqrt{n/B}$.
\end{proof}

We now pull everything together to bound the runtime of Type 2 HBP algorithms.

%\marginpar{RJC 3/7}
% RJC new theorem and corollary
\begin{theorem}
\label{thm:rws-run-times}
Let $\cal A$ be a limited-access,
top dominant Type 2 HBP algorithm with $S^l(n) = \Theta(n)$.
Let  $c \geq 1$ denote the number of collections of recursive calls made by $\cal A$.
Suppose that each recursive call $\cal{A}$ makes has size at most
$s(n) \le n/b$, for some constant $b > 1$.
Further suppose that $\cal{A}$ observes the constraints on writes for BP
and HBP computations as specified in this section, and let $\cal{A}$ have size $n$.
Let $W$ be the worst case operation count when $\cal A$ is executed sequentially,
$Q$ its worst case cache miss cost again when
executed sequentially, and let $C(S,n)$ be an upper bound on
the additional cache miss count when $\cal A$ incur $S$ steals in its execution.

Then, with probability $ 1 -2^{-\Theta(aT_{\infty})}$,
for integer $a = \omega(1)$, $\cal A$'s runtime is given by:
\[
O\left( \frac Wp + \frac{bQ}{p} + b\frac{C(S,n)}{p}  + \frac{S}{p}(s + bB) \right)
\]
and $S=O(O(p\cdot [1+a]h(t))$, where
$h(t) = O(T_{\infty} + \frac bs [\ell_2(t) + \ell_4(t)])$.
\end{theorem}
\begin{proof}
$\cal A$'s performance, in terms of operation count and cache miss costs is bounded by
the cost of a sequential computation plus the additional costs associated with the $S$
steals.
There are three components to these additional costs:
the additional cache misses;
the block wait cost, which by Lemma~\ref{lem:limited-write-intf-first} is $O(BS)$;
and the work performed in making successful and unsuccessful steals, which
by Theorem~\ref{thm:rws-work} is $O(p\cdot s[1+a]h(t))$ with probability $1-2^{-aT_{\infty}}$.
Further, $S=O(O(p\cdot [1+a]h(t))$ with probability $1-2^{-aT_{\infty}}$, again by
Theorem~\ref{thm:rws-work}.
Finally, $h(t)=\ell_1(t) + \frac bs [\ell_2(t) + \ell_3(t) + \ell_4(t)]
=O(T_{\infty} + \frac bs [\ell_2(t) + \ell_4(t)])$.
\end{proof}
Note that Lemma~\ref{lem:hbp-bound},
Corollary~\ref{cor:tighter-hbp-bound}, and Theorem~\ref{th:path-length}
provide bounds on
$\ell_2(t)$ and $\ell_4(t)$.

\begin{corollary}
\label{cor:opt-speedup}
Under the conditions of Theorem~\ref{thm:rws-run-times}, if $s=\Theta(b)$
and $C(S,n) + S \cdot B = O(Q)$ then the execution of $\cal A$ under RWS
using $p$ processors achieves an optimal $\Theta(p)$ speedup
compared to the sequential execution.
\end{corollary}

\section{Algorithms Runtimes}
\label{sec:alg-bounds}

%\marginpar{RJC 3/8}
% RJC moved and expanded

Bounds on the runtimes of the MM algorithms discussed earlier
can be readily deduced using Corollaries~\ref{cor:depth-n-MM-cache-miss}
and~\ref{cor:depth-log-2-n-MM-cache-miss},
Theorems~\ref{thm:rws-work} and~\ref{th:path-length},
and Corollary~\ref{cor:opt-speedup},
as shown in the next lemma.

\begin{lemma}
\label{lem:MM-runtime}
For each of the MM algorithms, let
$S$ be the number of steals it incurs.
In a sequential execution, each MM algorithm incurs $O(n^3/(BM^{1/2}))$
cache misses.
Under RWS, the steals cause an additional
$C(S,n)= O(n^3/(BM^{1/2}) + S^{1/3}\frac{n^2}{B} + S)$ cache misses.
The block delay is $O(S\cdot B)$.
This is optimal if $s=\Theta(b)$ and $S\cdot M^{1/2}\max\{B^2,M\} \le n^3$,
i.e.\ the runtime is
\[
O\left(\frac 1p \left[n^3 + b \frac{n^3}{B\cdot M^{1/2}} \right] \right).
\]

The depth $n$ matrix multiply algorithm, with local arrays for holding partial results, incurs,
with probability $1-1/2^{an}$, for $a = \omega(1)$,
$S= O([\frac{b+s}{s}pn + \frac bs pn\sqrt{B}][1+a])$ successful steals.
With $a=1$ and $s = \Theta(b)$, this is optimal if $p\le n^2/(B^{1/2}M^{3/2})$
% RJC 3/10
and $M \ge B^2$.

The depth $\log^2 n$ matrix multiply algorithm incurs,
with probability $1-1/2^{a\log^2 n}$, for $a = \omega(1)$,
$S= O([\frac{b+s}{s}p\log^2 n + \frac bs pB \log n][1+a])$ steals.
With $a=1$ and $s = \Theta(b)$, this is optimal if
$p(\log^2 n + B \log n) \le \frac{n^3}{M^{3/2}}$
% RJC 3/10
and $M \ge B^2$.
\end{lemma}
\begin{proof}
The bound on $C(S,n)$ is given in Corollaries~\ref{cor:depth-n-MM-cache-miss}
and~\ref{cor:depth-log-2-n-MM-cache-miss}.
By Corollary~\ref{cor:opt-speedup}, this is
optimal if (i)
$S\cdot B \le n^3/(BM^{1/2})$ or $S\cdot B^2 M^{1/2} \le n^3$, and (ii)
$S^{1/3}\frac{n^2}{B} \le n^3/(BM^{1/2})$ or $S\cdot M^{3/2} \le n^3$.

The depth $n$ algorithm has $T_{\infty}=O(n)$.
By Theorem~\ref{th:path-length},
$h(t) = O(n + \frac bs n\sqrt{B})$, as for this algorithm $s(n^2) = n^2/4$ and $c=2$.
Thus, by Theorem~\ref{thm:rws-work},
with probability $1-1/2^{an}$, for $a = \omega(1)$,
the depth $n$ algorithm incurs
$S= O([\frac{b+s}{s}pn + \frac bs pn\sqrt{B}][1+a])$ successful steals.
If $a=1$ and $s = \Theta(b)$, this is $O(pn\sqrt{B})$ steals.
On substituting for $S$, we see that this is optimal if
$p n \sqrt{B} \le \frac{n^3}{M^{3/2}}$ and $B^2 \le M$.

The depth $\log^2 n$ algorithm has $T_{\infty}=O(\log^2 n)$.
By Theorem~\ref{th:path-length},
$h(t) = O(\log^2 n + \frac bs B\log n)$, as for this algorithm
$s(n^2) = n^2/4$ but $c=1$.
Thus, by Theorem~\ref{thm:rws-work},
with probability $1-1/2^{a\log^2 n}$, for $a = \omega(1)$,
the depth $\log^2 n$ algorithm incurs
$S= O([\frac{b+s}{s}p\log^2 n + \frac bs pB\log n][1+a])$ successful steals.
If $a=1$ and $s = \Theta(b)$, this is $O(p\log n[\log n + B])$ steals.
On substituting for $S$, we see that this is optimal if
$p (\log^2 n + B\log n) \le \frac{n^3}{M^{3/2}}$ and $B^2 \le M$.
\end{proof}

Next, we state bounds for the following additional algorithms:
matrix transpose (when in BI format), FFT~\cite{CR10a}
and sorting~\cite{CR10b}.
(The FFT algorithm treats the data as being in a 2-D matrix, and repeatedly transposes
suitable submatrices. Again we assume the matrix is in BI format; as this algorithm
has $T_{\infty} = O(\log n\log\log n)$ a different algorithm for the BI to RM
conversion is needed. It is given in our companion paper~\cite{CR10a}.
Again, its costs are dominated in all regards by those for the FFT algorithm.)

\begin{theorem}
\label{cor:rws-run-times}
The following algorithms have the stated runtimes with probability $ 1 -2^{-\Theta(aT_{\infty})}$,
for integer $a\ = \omega(1)$, where $W$ is their operation count, $Q$ their cache miss count when
executed sequentially, and $C(S,n)$ their additional cache miss count when they incur $S$ steals.
\[
O\left( \frac Wp + \frac{bQ}{p} + b\frac{C(S,n)}{p}  + \frac{S}{p}(s + bB) \right).
\]

\vone
\noindent
\emph{(}i\emph{)} BP algorithms \emph{(}e.g.~prefix sums\emph{)}.
\\
$W= O(n)$, $Q=O(n/B)$, $T_{\infty} = O(\log n)$,
$S=O(p(\frac{b+s}{s}\log n + \frac bs B)(1+a))$, $C(S,n)=O(S)$.
Assuming that $s = \Theta(b)$ and $a = O(1)$,
this has combined cache and block miss costs of the same magnitude as the sequential cache miss cost
when $pB(\log n +  B) \le n$.

\vone
\noindent
\emph{(}ii\emph{)} Matrix transpose, RM to BI conversion.\\
The bounds from (i) apply, with $n^2$ replacing $n$, as this is a BP algorithm.
\emph{(}For matrix transpose, this is assuming that the matrix is in BI format.
Possible complementary BI to RM algorithms are discussed in~\cite{CR10a}.\emph{)}

\vone
\noindent
\emph{(}iii\emph{)} Sort:
See~\cite{CR10b} for a description of this algorithm.
\\
$W= O(n \log n)$, $Q= O(\frac nB \frac {\log n} {\log M})$ assuming that $M\ge B^2$
\emph{(}the ``tall cache assumption''\emph{)}, $T_{\infty} = O(\log n\log\log n)$,
$S=O(p(\frac{b+s}{s}\log n \log\log n +\frac bs  B \frac{\log n}{\log B})(1+a))$,
$C(S,n)=O(2^j\frac nB \frac {\log n} {\log M} )$
where $S =  O(\frac{2^jn}{M^{1/2^j}}\frac {\log n}{\log M})$,
for some integer $j \ge 1$.
Assuming that $s = \Theta(b)$ and $a = O(1)$,
this has combined cache and block miss costs of the same magnitude as the sequential cache miss cost
when $pB (\log n \log\log n + B \frac{\log n}{\log B}) \le \min\{\frac nB \frac {\log n} {\log M}, \frac{nB}{M^{1/2^j}}\frac {\log n}{\log M} \}$,
and $j=O(1)$, i.e.~$pB\log M ( \log\log n + B/\log B) \max\{B ,M^{1/2^j} \} \le n$.

\vone
\noindent
\emph{(}iv\emph{)} FFT:
See~\cite{CR10a} for a description of this algorithm.
The same bounds as for sorting apply.
\hide{
\vone
\noindent
(vi)
BI to gapped RM to RM:
\[
O\left( \frac{n^2}{p} + \frac{bn^2}{pB}
  + \left[  \log n + \frac{b}{s}B  \right]
                 \left[ b\frac{M}{B}+s+bB^2\log^2 B \right]
\right).
\]
The third term is bounded by $1/p$
times the sequential cache miss complexity, $O(n^2/B)$,
if $n^2 \ge p(\frac{M}{B} + B^2\log^2 B)(B +\log pM)$.

\vone
\noindent
(vii)
BI to RM(2), Method 1:
\[
O\left( \frac{n^2\log n}{p} + \frac{bn^2 \log n}{pB}
  + \left[ \log^2 n + \frac{b}{s}B\log n  \right] \left[ b\frac{M}{B}+s+bB \right]
\right).
\]
The third term is bounded by the second term
if $n^2\ge p (M+B^2)(B+ \log pM)$.
This cost is also upper bounded by the costs for the MM algorithms.

\vone
\noindent
(viii)
BI to RM(2), Method 2:
\[
O\left( \frac{n^2\log\log n}{p} + \frac{(i+1)bn^2 \log_B\log n}{pB}
  + \left[ \log n \log\log n + \frac{b}{s}B\log_B\log n  \right] \left[ b\frac{M^{1/2^i}}{B}+s+bB \right]
\right),
\]
for any fixed integer $i\ge 0$.
The third term is bounded by the second term
if $n^2 \ge p(\sqrt{M} + B^2)(B + \log pM)$,
on taking $i=1$.
This cost is also upper bounded by the costs for the FFT algorithm.
} % end hide

\hide{
\vone
\noindent
(v) List ranking:
\[
O\left( \frac{n\log n}{p} + \frac{jn}{B}\frac{\log n}{\log M}
  + \left[ \log^2 n \log\log n
  + \frac{b}{s} B \frac{\log^2 n} {\log B} \right]
    \left[ b\frac{M^{1/j}}{B} + s_R + bB \right] \right).
\]
The third term is bounded by $1/p$
times the sequential cache miss complexity,
$O(\frac{n}{B}\frac{\log n}{\log M})$,
if $M\ge B^2$ and
$n/log n \ge  \Theta(p[\sqrt{M} + B^2] \log M [\frac{B}{\log B} + \log\log pM])$
(again taking $j=2$).
} % end hide
\end{theorem}
\begin{proof}
%\marginpar{RJC 3/10}
% RJC 3new sentence
The main bound as a function of $S$ is given by Theorem~\ref{thm:rws-run-times}.

We explain the sorting bounds
%RJC 3/10
in more detail.
A task of size $r$ incurs $O(r/B + \sqrt{r})= O(r/B +B)$ cache misses in this algorithm.
As the block miss cost of a stolen task is bounded by $O(B)$, the second term in the
cache miss bound is absorbed into the block miss cost for the purposes of the analysis.

Here, there are collections of recursive problems of sizes $n$, $\sqrt{n}$, $n^{\frac14}$, etc.
There are 2 collections of the problems of size $\sqrt{n}$, 4 of those of size $n^{\frac14}$, etc.
For each collection, there are $\Theta(n/B)$ cache misses, when the problems are of size $2M$ or larger.
For subproblems of size $M$ or smaller, steals may result in another $O(n/B)$ cache misses
for each full collection of subproblems.
As usual, the worst case arises if the largest possible size subproblems are stolen.
This occurs with $j$ chosen as small as possible so that
$S= O(\frac{2^j n}{M^{1/2^j}}\frac {\log n}{\log M})$.

The block miss cost is $O(S \cdot B)$.
This yields the two bounds on $p$ in the final result.
\end{proof}

The list ranking algorithm in~\cite{CR10a} iterates a sorting algorithm $O(\log n)$
times, so its bounds are no more than $O(\log n)$ times the bounds for the sort
given above (in fact, somewhat better bounds can be obtained, as the combined
size of the sorting problems is $O(n\log n)$).

The connected components algorithm in~\cite{CR10a} iterates the list ranking
algorithm $O(\log n)$ times, so its bounds are no more than
$O(\log n)$ times the bounds for the list ranking algorithm,
but the sizes of each successive list ranking problem are the same.

% end hide
\noindent

\hide{
\vone
\noindent
(iii) Matrix multiply using 8-way recursion:
\\
$W= O(n^3)$, $Q=O(n^3/(B\sqrt{M}))$, $S=O(p(\frac{b+s}{s}\log^2 n + \frac bs B\log n)(1+a))$, $C(S,n)=O(S+ S^{1/3}n^2/B)$.
Assuming that $s = \Theta(b)$ and $a = O(1)$,
this has combined cache and block miss costs of the same magnitude as the sequential cache miss cost
when $p (\log^2 n + B \log n) \le \frac{n^3}{M^{3/2}}$.

\vone
\noindent
(iv)
$O(n)$-depth
matrix multiply.
\\
$W= O(n^3)$, $Q=O(n^3/(B\sqrt{M}))$, $S=O(p(\frac{b+s}{s}n + \frac bs n\sqrt{B})(1+a))$, $C(S,n)=O(S+ S^{1/3}n^2/B)$.
Assuming that $s = \Theta(b)$ and $a = O(1)$,
this has combined cache and block miss costs of the same magnitude as the sequential cache miss cost
when $p  \le \frac{n^2}{B^{1/2}M^{3/2}}$.
} % end hide

\bibliographystyle{abbrv}
\bibliography{sort,rws-refs}

\end{document}